\documentclass[aps,bibnotes,pra,nofootinbib]{revtex4}
\usepackage{youngtab,pstricks}
\usepackage{amsmath,amsfonts,amssymb}
\usepackage{graphicx}
\usepackage[frame,line,arrow,matrix,tips]{xy}
\CompilePrefix{xygui-}

\newtheorem{theorem}{Theorem}

\def\bmath#1{\mbox{\boldmath$#1$}}

\def\ra{\rightarrow}
\def\da{{\!\downarrow}}

\def\l{\left}
\def\r{\right}
\def\be{\begin{equation}}
\def\ee{\end{equation}}
\def\bea{\begin{eqnarray}}
\def\eea{\end{eqnarray}}
\newcommand{\bra}[1]{\mbox{$\left\langle #1 \right|$}}
\newcommand{\ket}[1]{\mbox{$\left| #1 \right\rangle$}}

\newcommand{\oprod}[1]{\mbox{\ket{#1}\bra{#1}}}
\newcommand{\eq}[1]{Eq.~(\ref{eq:#1})}
\newcommand{\eqs}[2]{Eqns.~(\ref{eq:#1}) and (\ref{eq:#2})}

\newcommand{\sect}[1]{Sec.~\ref{sec:#1}}
\newcommand{\fig}[1]{Fig.~\ref{fig:#1}}
\newcommand{\mscite}[1]{Ref.~\cite{#1}}

\def\>{\rangle}
\def\<{\langle}
\def\ot{\otimes}

\def\bp{{{\bf p}}}
\def\bq{{{\bf q}}}
\def\br{{{\bf r}}}
\def\bI{{{\bf I}}}
\def\bP{{{\bf P}}}
\def\bQ{{{\bf Q}}}
\def\bR{{{\bf R}}}
\def\bT{{{\bf T}}}

\def\cA{{\cal A}}
\def\cB{{\cal B}}
\def\cG{{\cal G}}
\def\cH{{\cal H}}
\def\cI{{\cal I}}

\def\cP{{\cal P}}
\def\cQ{{\cal Q}}
\def\cS{{\cal S}}
\def\cU{{\cal U}}
\def\bbC{\mathbb{C}}
\def\bbZ{\mathbb{Z}}

\def\Usch{{\bf U}_{\text{Sch}}}
\def\Ucg{{\bf U}_{\text{CG}}}
\def\Cdn{(\bbC^d)^{\ot n}}
\def\eps{\epsilon}

\def\st{\text{~s.t.~}}
\newcommand{\iso}[1]{\stackrel{#1}{\cong}}

\def\interlaces{{\precsim}}
\def\reverselaces{{\succsim}}

\DeclareMathOperator{\End}{End}
\DeclareMathOperator{\Hom}{Hom}
\DeclareMathOperator{\poly}{poly}
\DeclareMathOperator{\sgn}{sgn}
\DeclareMathOperator{\Span}{Span}

\def\w{\ar@{-}[l]}

\def\nw{*-{~}\w}

\def\b{*={\bullet}}

\def\op#1{*+[F]{\rule[-0.2ex]{0ex}{2.1ex}#1}}   

\newbox{\ssbox}

\def\gspace#1{*+{\rule[-0.2ex]{0ex}{2.1ex}%
    \setbox\ssbox=\hbox{$#1$}%
    \hspace*{\wd\ssbox}}}

\def\gnqubit#1#2{\gspace{#1}
         \save [].[#2]!C="qq"*[F]\frm{}\restore
         \save "qq"*[]{#1} \restore}

\def\RCSid $#1${\def\rcsid{#1}}
\RCSid $Id: longschur.tex,v 1.18 2005/12/03 21:03:21 aram Exp $

\def\RCSSplitField #1: #2 #3 {\def\RCSdate{#2}\def\RCStime{#3}}
\def\RCS $#1${\RCSSplitField #1}\RCS $Date: 2005/12/03 21:03:21 $

\begin{document}

\title{The Quantum Schur Transform:\\ I. Efficient Qudit Circuits}
\author{D. Bacon}
\affiliation{Dept.~of Computer Science and Engineering, Univ.~of
Washington, Seattle, WA, USA}
\affiliation{Santa Fe Institute, Santa Fe, NM 87501 USA}
\affiliation{Institute for Quantum Information, California Institute of
Technology, Pasadena, CA, USA} 
\affiliation{Dept.~of Physics, California Institute of
Technology, Pasadena, CA, USA}
\author{I.~Chuang}
\affiliation{Center for Bits and Atoms, Massachusetts Institute of Technology,
Cambridge, MA, USA} 
\affiliation{Dept.~of Electrical Engineering and Computer Science,
Massachusetts Institute of 
Technology, Cambridge, MA, USA}
\affiliation{Dept.~of Physics, Massachusetts Institute of
Technology, Cambridge, MA, USA}
\author{A.~Harrow}
\affiliation{Center for Bits and Atoms, Massachusetts Institute of Technology,
Cambridge, MA, USA} 
\affiliation{Dept.~of Physics, Massachusetts Institute of
Technology, Cambridge, MA, USA}
\affiliation{Dept.~of Computer Science, Univ.~of Bristol,
Bristol, U.K.}
\date{\today}
\pacs{03.67.-a,03.67.Lx,03.67.Mn}

\begin{abstract}
We present an efficient family of quantum circuits for a fundamental primitive
in quantum information theory, the Schur transform.  The Schur transform on $n$
$d$ dimensional quantum systems is a transform between a standard computational
basis to a labelling related to the representation theory of the symmetric and
unitary groups.  If we desire to implement the Schur transform to an accuracy
of $\epsilon$, then our circuit construction uses a number of gates which is
polynomial in $n$, $d$ and $\log(\epsilon^{-1})$.  The important insights we
use to perform this construction are the selection of the appropriate subgroup
adapted basis and the Wigner-Eckart theorem.  Our efficient circuit
construction renders numerous protocols in quantum information theory
computationally tractable and is an important new efficient quantum circuit
family which goes significantly beyond the standard paradigm of the quantum
Fourier transform.
\end{abstract}


\maketitle

\section{Introduction}

The last decade has seen the development and expansion of a robust theory of
quantum
information\cite{Bennett:92a,Bennett:93a,Schumacher:95a,Bennett:96a,Bennett:96b,Devetak:03a}.
The basic goal of this new work has been the identification and quantification
of different information resources in situations where the laws of quantum
theory are applied to the physical carriers of information. Quantum information
theory has made great progress in understanding the optimal rates of the
manipulation and transmission of quantum information.  Despite this success,
however, much of the work in quantum information theory may not be of practical
value.  This is because most of the work in quantum information theory has
focused on protocols which allow for unbounded quantum computational resources.
Thus while the transforms in the quantum information protocols are well
defined, whether these transforms can be implemented with quantum circuits
whose size scales efficiently with the size of the quantum information problem
is often left unaddressed.  An analogous situation arises classically, for
example, in the theory of classical error correcting codes.  On the one hand,
we would like the classical error correcting code to attain some characteristic
efficiency for communicating over a noisy channel.  On the other hand, we would
also like to design codes whose encoding and decoding does not significantly
lag our communication. In order to be of practical value a classical error
correcting code must use computational resources which scale at a reasonable
rate.  While the goal of performing classical coding tasks in polynomial or
even linear time has long been studied, quantum information theory results have
typically ignored questions of efficiency. For example, random quantum coding
results (such as \cite{Holevo98,SW97,BHLSW03,DW03c}) require an exponential
number of bits to describe, and like classical random coding techniques, do not
yield efficient algorithms. There are a few important exceptions.  Some quantum
coding tasks, such as Schumacher compression\cite{Schumacher:95a,JS94}, are
essentially equivalent to classical circuits, and as such can be performed
efficiently on a quantum computer by carefully modifying an efficient classical
algorithm to run reversibly and to deal properly with ancilla
systems\cite{Cleve:96a}.  Another example, which illustrates some of the
challenges involved, is \mscite{Kaye01}'s efficient implementation of
entanglement concentration\cite{Bennett:96b}.  Quantum key
distribution\cite{Bennett:84a} not only runs efficiently, but can be
implemented with entirely, or almost entirely, single-qubit operations and
classical computation.  Fault tolerant quantum computing\cite{Shor:96a} usually
seeks to perform error correction with as few gates as possible, although with
teleportation-based techniques\cite{Gottesman:99a,Knill:04a} computational
efficiency may not be quite as critical to the threshold rate.  Finally, some
randomized quantum code constructions have been given efficient constructions
using classical derandomization techniques in \cite{AS04}.  In this paper we
present an efficient family of quantum circuits for a transform used
ubiquitously\cite{Keyl:01a,Gill:02a,Vidal:99a,Hayashi:02a,Hayashi:02b,Hayashi:02c,Hayashi:02d,Zanardi:97a,Knill:00a,Kempe:01a,Bacon:01a,Bartlett:03a}
in quantum information protocols, the Schur transform.  Our efficient
construction of the Schur transform adds to the above list a powerful new tool
for finding algorithms that implement quantum communication tasks.

The Schur transform is a unitary transform on $n$ $d$-dimensional quantum
systems ($n$ qudits). The basis change corresponding to the Schur transform
goes from a standard computational basis on the $n$ qudits to a labelling
related to the representation theory of the symmetric and unitary groups; much
like the Fourier transform, it thus transforms from a local to a more global,
collective basis, which captures symmetries of the system.  In this article we
show how to efficiently implement the Schur transform as a quantum circuit. The
size of the circuit we construct is polynomial in the number of qudits, $n$,
the dimension of the individual quantum systems, $d$, and the log of accuracy
to which we implement the transform, $\log(\epsilon^{-1})$.  Our efficient
quantum circuit for the Schur transform makes possible efficient quantum
circuits for numerous quantum information tasks: optimal spectrum
estimation\cite{Keyl:01a,Gill:02a}, universal entanglement
concentration\cite{Hayashi:02a}, universal compression with optimal overflow
exponent\cite{Hayashi:02b,Hayashi:02c}, encoding into decoherence-free
subsystems\cite{Zanardi:97a,Knill:00a,Kempe:01a,Bacon:01a}, optimal hypothesis
testing\cite{Hayashi:02d}, and quantum and classical communication without
shared reference frames\cite{Bartlett:03a}.
The central role of the Schur transform in all of these protocols is
due to the fact that the symmetries of independent and identically
distributed quantum
states are naturally treated by the representation theory of the
symmetric and unitary groups.
Thus in addition to making practical these
quantum information protocols, the Schur transform is an interesting new
unitary transformation with an interpretation relating to these symmetries.

There are many difficulties in designing an efficient quantum circuit for the
Schur transform which we overcome in this paper.  The first difficulty is in
efficiently representing the basis used in the Schur transform, the Schur
basis.  A second difficulty comes in the actual circuit construction.  In
particular we would like to construct the Schur transform from a series of
Clebsch-Gordan transforms.  However, it is not at all obvious how to
efficiently implement these Clebsch-Gordan transforms, nor is obvious that such
a cascade can perform the complete Schur transform.

Our resolution to these problems begins by our selection of certain
subgroup-adapted bases for the Schur basis.  In particular we use the
Gel'fand-Zetlin basis\cite{Gelfand:50a} and the Young-Yamanouchi basis
(sometimes called Young's orthogonal basis)\cite{James:81a}.  We note that
subgroup-adapted bases are also used in constructing efficient quantum circuits
for Fourier transforms over nonabelian finite groups\cite{Moore:03a}. However,
we should emphasize that the Schur transform is not a Fourier
transform over a finite group, although connections between such transforms and
the Schur transform exist, and are discussed in part II of this paper.  By
choosing the Gel'fand-Zetlin basis and the Young-Yamanouchi basis, we are able
to show that the Schur transform can be constructed from a cascade of
Clebsch-Gordan transforms.  Further, the use of the Gel'fand-Zetlin basis,
combined with the Wigner-Eckart theorem, allows us to efficiently implement the
Clebsch-Gordan transform.  In particular the Wigner-Eckart theorem allows us to
recursively express the $d$ dimensional Clebsch-Gordan transform in terms of
the $d-1$ dimensional Clebsch-Gordan transform and small, efficiently
implementable, unitary transforms.  This produces an efficient recursive
construction of the Clebsch-Gordan transform.  Without the recursive structure
we exploit, a naive circuit construction would seem to require $n^{O(d^2)}$
gates.  Our recursive exploitation of the Wigner-Eckart theorem allows us to
implement the Clebsch-Gordan transform to accuracy $\epsilon$ using
$\poly(d,\log n, \log 1/\epsilon)$ gates.  The total size of our circuit
construction for the Schur transform is $n \poly (d,\log n, \log 1/\epsilon)$.

The outline of the paper is as follows.  In Section \ref{sec:schur} we
introduce the Schur transform, along with basic concepts from
representation theory, and review the numerous applications of the Schur
transform in quantum information theory.  In Section \ref{sec:gz} we introduce
the basis labelling scheme used in the Schur transformation using the
concept of a subgroup-adapted basis.  Once we have a concrete Schur
basis defined, we describe the Clebsch-Gordan transform and explain
how to use it to give an efficient circuit for the Schur transform in
\sect{schur-circuit}.  Finally, we complete the algorithm in
\sect{cg-construct} by
constructing an efficient circuit for the Clebsch-Gordan transform.

\section{The Schur Transform and Its Applications} \label{sec:schur}

Consider a system of $n$ $d$-dimensional quantum systems: $n$
qudits.  Fix a standard computational basis $|i\rangle$, $i=1\dots
d$ for the state space of each qudit: $\bbC^d$.  A basis for the
system $(\bbC^d)^{\ot n}$ is then $|i_1\rangle
\otimes |i_2\rangle \otimes \cdots \otimes
|i_n\rangle=|i_1,i_2,\dots,i_n\rangle$ where $i_k=1\dots d$.  The
Schur transform is a unitary transform on the standard basis
$|i_1,i_2,\dots,i_n\rangle$.  After the Schur transform, the
standard computational basis is relabeled as $\ket{\lambda}
\ket{p}\ket{q}$ (symbols which we later define).  In this section we
review the basic
representation theory necessary to understand the Schur basis and
also review the applications of this transformation to different
protocols in quantum information theory.


\subsection{Representation theory background}

The Schur transform is related to the representations of two groups on
$(\bbC^d)^{\ot n}$, a representation of the symmetric group and
a representation of the unitary group.  We first recall the
basics of representation theory before introducing these
representations.
For a more detailed description of
representation theory, the reader should consult \cite{Artin95} for
general facts about group theory and representation theory or
\cite{Goodman:98a} for representations of Lie groups.  See also
\cite{Fulton91} for a more introductory and informal approach to Lie
groups and their representations.

{\em Representations:} For a complex vector space $V$, define
$\End(V)$ to be set of linear maps from $V$ to itself (endomorphisms).
 A representation of a group ${\mathcal G}$ is a vector
space $V$ together with a homomorphism from $\cG$ to $\End(V)$, i.e. a function
${\bf R}:\cG\ra \End(V)$ such that ${\bf R}(g_1) {\bf R}(g_2)= {\bf R}(g_1
g_2)$.  If ${\bf R}(g)$ is a unitary operator for all $g$, then we say ${\bf
R}$ is a unitary representation.  Furthermore, we say a representation
$(\bR,V)$ is finite dimensional if $V$ is a finite dimensional vector space. In
this paper, we will always consider complex finite dimensional, unitary
representations and use the generic term `representation' to refer to complex,
finite dimensional, unitary representations.  Also, when clear from the
context, we will denote a representation $(\bR,V)$ simply by the representation
space $V$.

The reason we consider only complex, finite dimensional, unitary
representations is so that we can use them in quantum computing.  If
$d=\dim V$, then a $d$-dimensional quantum system can hold a unit
vector in a representation $V$.  A group element $g\in\cG$ corresponds
to a unitary rotation $\bR(g)$, which can in principle be performed by
a quantum computer.

{\em Homomorphisms:} For any two vector spaces $V_1$ and $V_2$, define
$\Hom(V_1,V_2)$ to be the set of linear transformations from $V_1$ to $V_2$. If
$\cG$ acts on $V_1$ and $V_2$ with representation matrices $\bR_1$ and $\bR_2$
respectively, then the canonical action of $\cG$ on $\Hom(V_1,V_2)$ is given by
the map from $M$ to $\bR_2(g)M\bR_1(g)^{-1}$ for any $M\in\Hom(V_1,V_2)$.  For
any representation $(\bR,V)$ define $V^\cG$ to be the space of $\cG$-invariant
vectors of $V$: i.e. $V^{\cG}:=\{\ket{v}\in V : \bR(g)\ket{v}=\ket{v} \,\forall
g\in\cG\}$.  Of particular interest is the space $\Hom(V_1,V_2)^\cG$, which can
be thought of as the linear maps from $V_1$ to $V_2$ which commute with the
action of $\cG$. If $\Hom(V_1,V_2)^\cG$ contains any invertible maps (or
equivalently, any unitary maps) then we say that $(\bR_1,V_1)$ and
$(\bR_2,V_2)$ are {\em equivalent} representations and write
$$V_1 \stackrel{\cG}{\cong} V_2.$$
This means that there exists a unitary change of basis $U:V_1\ra V_2$ such
that for any $g\in\cG$, $U\bR_1(g)U^\dag = \bR_2(g)$.

{\em Dual representations:} Recall that the {\em dual} of a vector
space $V$ is the set of linear maps from $V$ to $\bbC$ and is denoted
$V^*$.  Usually if vectors in $V$ are denoted by kets (e.g. $\ket{v}$)
then vectors in $V^*$ are denoted by bras (e.g. $\bra{v}$).  If we fix
a basis $\{\ket{v_1},\ket{v_2},\ldots\}$ for $V$ then the transpose is
a linear map from $V$ to $V^*$ given by $\ket{v_i}\ra \bra{v_i}$.
Now, for a representation $(\bR,V)$ we can define the {\em dual
representation} $(\bR^*,V^*)$ by $\bR^*(g)\bra{v^*} := \bra{v^*}
\bR(g^{-1})$.  If we think of $\bR^*$ as a representation on $V$
(using the transpose map to relate $V$ and $V^*$), then
it is given by $\bR^*(g)=(\bR(g^{-1}))^T$.  When $\bR$ is a unitary
representation, this is the same as the {\em conjugate representation}
$\bR(g)^*$, where here $^*$ denotes the entrywise complex conjugate.
One can readily verify that the dual and conjugate representations are
indeed representations and that
$\Hom(V_1,V_2)\stackrel{\cG}{\cong}V_1^* \ot V_2$.

{\em Irreducible representations:} Generically the unitary operators
of a representation may be specified (and manipulated on a quantum
computer) in an arbitrary orthonormal basis.  The added structure of
being a representation, however, implies that there are particular
bases which are more fundamental to expressing the action of the
group.  We say a representation $(\bR,V)$ is irreducible (and call it
an irreducible representaiton, or {\em irrep}) if the only subspaces
of $V$ which are invariant under $\bR$ are the empty subspace $\{0\}$
and the entire space $V$.  For finite groups, any finite-dimensional
complex representation is reducible; meaning it is decomposable into a
direct sum of irreps.  For Lie groups, we need additional conditions,
such as demanding that the representation $\bR(g)$ be {\em rational};
i.e. its matrix elements are polynomial functions of the matrix
elements $g_{ij}$ and $(\det g)^{-1}$.  We say a representation of a
Lie group is {\em polynomial} if its matrix elements are polynomial
functions only of the $g_{ij}$.

{\em Isotypic decomposition:} Let $\hat{\cG}$ be a complete set of
inequivalent irreps of $\cG$.  Then for any reducible representation
$({\bf R},V)$ there is a basis under which the action of ${\bf R}(g)$
can be expressed as
\begin{equation}
{\bf R}(g)\cong \bigoplus_{\lambda\in\hat{\cG}}
\bigoplus_{j=1}^{n_\lambda}{\bf r}_\lambda(g)=
\bigoplus_{\lambda\in\hat{\cG}}  {\br}_\lambda(g)
\ot {\bf I}_{n_\lambda}  \label{eq:directsum}
\end{equation}
where $\lambda\in \hat{\cG}$ labels an irrep $({\bf
r}_\lambda,V_\lambda)$ and $n_\lambda$ is the multiplicity of the
irrep $\lambda$ in the representation $V$.  Here we use $\cong$ to
indicate that there exists a unitary change of basis relating the
left-hand size to the right-hand side.\footnote{We only need to use
$\iso{\cG}$ when relating representation spaces.  In \eq{directsum} and
other similar isomorphisms, we instead explicitly specify the
dependence of both sides on $g\in\cG$.}  Under this change of basis we
obtain a similar decomposition of the representation space $V$ (known
as the {\em isotypic decomposition}):
\be V \stackrel{\cG}{\cong} \bigoplus_{\lambda\in\hat{\cG}}
 V_\lambda \ot \bbC^{n_\lambda}.
\label{eq:rep-space-decomp}\ee
Thus while generically we may be given a representation in some
arbitrary basis, the structure of being a representation picks out a
particular basis under which the action of the representation is not
just block diagonal but also maximally block diagonal: a direct sum of
irreps.

Moreover, the multiplicity space $\bbC^{n_\lambda}$
in \eq{rep-space-decomp} has the structure of $\Hom(V_\lambda,V)^\cG$.
This means that for any representation $(\bR,V)$,
\eq{rep-space-decomp} can be restated as
\be V \stackrel{\cG}{\cong}
 \bigoplus_{\lambda\in\hat{\cG}} V_\lambda \ot \Hom(V_\lambda, V)^{\cG}.
 \label{eq:hom-decomp}\ee
Since $\cG$ acts trivially on $\Hom(V_\lambda, V)^{\cG}$,
\eq{directsum} remains the same.  As with the other results in this
chapter, a proof of \eq{hom-decomp} can be found in
\cite{Goodman:98a}, or other standard texts on representation theory.

The value of \eq{hom-decomp} is that the unitary mapping from the
right-hand side (RHS)
to the left-hand side (LHS) has a simple explicit expression: it
corresponds to the canonical map $\varphi:A\otimes \Hom(A,B)\ra B$ given by
$\varphi(a\otimes f)=f(a)$.  Of course, this doesn't tell us how to
describe $\Hom(V_\lambda,V)^{\cG}$, or how to specify an orthonormal
basis for the space, but we will later find this form of the
decomposition useful.

\subsection{The Schur Transform}

We now turn to the two representations relevant to the Schur transform.  Recall
that the symmetric group ${\mathcal S}_n$ of degree $n$, is the group of all
permutations of $n$ objects. Then we have the following natural representation
of the symmetric group on the space $(\bbC^d)^{\ot n}$:
\begin{equation}
\bP(s) |i_1\rangle \otimes |i_2\rangle \otimes \cdots
\otimes|i_n\rangle = |i_{s^{-1}(1)}\rangle\otimes
|i_{s^{-1}(2)}\rangle\otimes \cdots
\otimes|i_{s^{-1}(n)}\rangle
\end{equation}
where $s \in {\mathcal S}_n$ is a permutation and $s(i)$ is the label
describing the action of $s$ on label $i$.  For example, if we are
considering ${\mathcal S}_3$ and the permutation we are considering is the
transposition $s=(12)$, then $\bP(s)
|i_1,i_2,i_3\rangle=|i_2,i_1,i_3\rangle$.  $(\bP,\Cdn)$ is the representation
of the symmetric group which will be relevant to the Schur transform.
Note that $\bP$ obviously depends on $n$, but also has an implicit
dependence on $d$.

Now we
turn to the representation of the unitary group.  Let ${\mathcal U}_d$ denote
the group of $d\times d$ unitary operators. Then there is a representation of
$\cU_d$ given by the $n$-fold product action as
\begin{equation}
{\bf Q}(U)|i_1\rangle \otimes|i_2\rangle \otimes\cdots\otimes |i_n\rangle =
U |i_1\rangle\otimes U|i_2\rangle \otimes\cdots \otimes U|i_n\rangle
\end{equation}
for any $U \in {\mathcal U}_d$. More compactly, we could write that
$\bQ(U)=U^{\ot n}$.  $({\bf Q}, \Cdn)$ is the representation of the
unitary group
which will be relevant to the Schur transform.

Since both $\bP(s)$ and ${\bf Q}(U)$  meet our above criteria for
reducibility, they can be decomposed into a direct sum of irreps as in
\eq{directsum},
\begin{eqnarray}
\bP(s)&\stackrel{\cS_n}{\cong}
&\bigoplus_\alpha {\bf I}_{n_\alpha} \otimes
{\bf p}_\alpha(s)
\nonumber \\
{\bf Q}(U)& \stackrel{\cU_d}{\cong}
&\bigoplus_\beta {\bf I}_{m_\beta} \otimes {\bf
q}_\beta(U) \label{eq:fistdecomp}
\end{eqnarray}
where $n_\alpha$ ($m_\beta$) is the multiplicity of the $\alpha$th
($\beta$th) irrep ${\bf p}_\alpha(s)$ (${\bf q}_\beta(U)$) in
the representation $\bP(s)$ (${\bf Q}(U)$).  At this point there is
not necessarily any relation between the two different unitary
transforms implementing the isomorphisms in \eq{fistdecomp}.  However,
further structure in this
decomposition follows from the fact that $\bP(s)$
commutes with ${\bf Q}(U)$: $\bP(s) {\bf Q}(U)={\bf Q}(U)
\bP(s)$.  This implies, via Schur's Lemma, that the action
of the irreps of $\bP(s)$ must act on the multiplicity
labels of the irreps ${\bf Q}(U)$ and vice versa.  Thus, the
simultaneous action of $\bP$ and $\bQ$ on $(\bbC^d)^{\ot n}$
decomposes as
\begin{equation}
 \bQ(U) \bP(s)\stackrel{\cU_d\times\cS_n}{\cong}
 \bigoplus_\alpha\bigoplus_\beta
 {\bf I}_{m_{\alpha,\beta}} \otimes
\bq_\beta(U) \ot \bp_\alpha(s)
\label{eq:pqdecomp}\end{equation}
where $m_{\alpha,\beta}$ can be thought of as the multiplicity of the
irrep ${\bf p}_\alpha(s) \ot \bq_\beta(U)$ of the group
$\cU_d\times\cS_n$.

Not only do $\bP$ and $\bQ$ commute, but the algebras they generate
(i.e. ${\cal A} := \bP(\bbC[\cS_n]) = \Span\{\bP(s) : s\in\cS_n\}$
and ${\cal B} := \bQ(\bbC[\cU_d]) = \Span\{\bQ(U) : U\in\cU_d\}$) {\em
centralize} each other\cite{Goodman:98a}, meaning that $\cB$ is the
set of operators in
$\End((\bbC^d)^{\ot n})$ commuting with $\cA$ and vice versa, $\cA$ is
the set of operators in
$\End((\bbC^d)^{\ot n})$ commuting with $\cB$.  This means that the
multiplicities $m_{\alpha,\beta}$ are either zero or one, and that
each $\alpha$ and $\beta$ appears at most once.  Thus \eq{pqdecomp}
can be further simplified to
\be \bQ(U)\bP(s) \stackrel{\cU_d\times\cS_n}{\cong}
 \bigoplus_\lambda   {\bf q}_\lambda(U)
\ot \bp_\lambda(s)\label{eq:preschur-decomp}\ee
where $\lambda$ runs over some unspecified set.

Finally, Schur duality (or Schur-Weyl
duality)\cite{Goodman:98a,Chen:02a} provides a simple
characterization of the range of $\lambda$ in \eq{preschur-decomp} and
shows how the decompositions are related for different values of $n$
and $d$.  To define Schur duality, we will first need to specify the
irreps of $\cS_n$ and $\cU_d$.

Let $\mathcal{I}_{d,n}=\{\lambda=(\lambda_1,
\lambda_2,\dots,\lambda_d) |\lambda_1 \geq \lambda _2 \geq \cdots
\geq \lambda_d \geq 0$ and $\sum_{i=1}^d \lambda_i = n\}$ denote
partitions of $n$ into $\leq d$ parts.  We consider two partitions
$(\lambda_1,\ldots,\lambda_d)$ and
$(\lambda_1,\ldots,\lambda_d,0,\ldots,0)$
equivalent if they differ only by trailing zeroes; according to this
principle, $\cI_n := \cI_{n,n}$ contains all the partitions of $n$.
Partitions label irreps of $\cS_n$ and $\cU_d$ as follows: if we let $d$
vary, then $\cI_{d,n}$ labels irreps of $\cS_n$, and if we let $n$
vary, then $\cI_{d,n}$ labels polynomial irreps of $\cU_d$. Call these
$({\bf p}_\lambda,\cP_\lambda)$ and $({\bf q}_\lambda^d,
\cQ_\lambda^d)$ respectively, for $\lambda\in\cI_{d,n}$.
We need the superscript $d$ because the same partition $\lambda$ can
label different irreps for different $\cU_d$; on the other hand the
$\cS_n$-irrep
$\cP_\lambda$ is uniquely labeled by $\lambda$ since $n=\sum_i \lambda_i$.

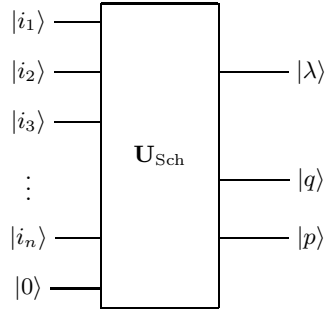
\begin{figure}[ht]
\begin{centering}
\leavevmode\xymatrix@R=4pt@C=4pt{
{\ket{i_1}} & *{~~~} & \gnqubit{{~~~\Usch~~~}}{ddddd}\ar@{-}[ll]\\
{\ket{i_2}} & \nw & \gspace{~~~\Usch~~~}\w && |\lambda\>\ar@{-}[ll] \\
{\ket{i_3}} & \nw &\gspace{~~~\Usch~~~}\w & ~~~~\\
{\vdots} &  & \gspace{~~~\Usch~~~} && |q\>\ar@{-}[ll]\\
{\ket{i_n}} & \nw & \gspace{~~~\Usch~~~}\w && |p\>\ar@{-}[ll]\\
{\ket{0}}& \nw & \gspace{~~~\Usch~~~}\w
}
\caption{The Schur transform.  Notice how the direct sum over
  $\lambda$ in \eq{schur-decomp} becomes a tensor product between
the $\ket{\lambda}$ register and the $\ket{q}$ and $\ket{p}$
registers.  Since the number of qubits needed for $\ket{q}$ and
$\ket{p}$ vary with $\lambda$, we need slightly more spatial
resources, which are here denoted by the ancilla input $|0\rangle$.}
\label{fig:schur}
\end{centering}
\end{figure}

For the case of $n$ qudits, Schur
duality states that there exists a basis (which we label
$\ket{\lambda}\ket{q_\lambda}\ket{p_\lambda}_{\rm Sch}$ and call the
{\em Schur basis}) which simultaneously decomposes the action of
$\bP(s)$ and $\bQ(U)$ into irreps:
\begin{eqnarray}
{\bf Q}(U)\ket{\lambda}\ket{q_\lambda}\ket{p_\lambda}_{\rm Sch}&=&
\ket{\lambda}({\bf q}_\lambda^d(U) \ket{q_\lambda})
\ket{p_\lambda}_{\rm Sch}
\nonumber \\
\bP(s) \ket{\lambda}\ket{q_\lambda}\ket{p_\lambda}_{\rm Sch}
&=& \ket{\lambda} \ket{q_\lambda} ({\bf
p}_\lambda(s)\ket{p_\lambda})_{\rm Sch} \label{eq:schur}
\end{eqnarray}
and that the common representation space $(\bbC^d)^{\ot n}$ decomposes as
\be (\bbC^d)^{\ot n} \stackrel{\cU_d\times\cS_n}{\cong}
\bigoplus_{\lambda\in\cI_{d,n}} \cQ_\lambda^d \ot \cP_\lambda.
\label{eq:schur-decomp}\ee
The Schur basis can be expressed as superpositions over the
standard computational basis states $|i_1,i_2,\dots,i_n\rangle$ as
\begin{equation}
|\lambda,q_\lambda,p_\lambda\rangle_{\rm
Sch}=\sum_{i_1,i_2,\dots,i_n} \left[ {\bf U}_{\rm Sch}
\right]^{\lambda,q_\lambda,p_\lambda}_{i_1,i_2,\dots,i_n}
|i_1i_2\dots i_n \rangle,
\label{eq:Usch-def}\end{equation}
where ${\bf U}_{\rm Sch}$ is the unitary transformation implementing
the isomorphism in \eq{schur-decomp}.  Thus, for any $U\in\cU_d$ and
any $s\in\cS_n$,
\be \Usch \bQ(U)\bP(s) \Usch^\dag =
\sum_{\lambda\in\cI_{d,n}}
\oprod{\lambda} \ot \bq_\lambda^d(U) \ot \bp_\lambda(s).
\label{eq:schur-decomp2}\ee
If we now think of ${\bf U}_{\rm Sch}$ as a quantum circuit, it will
map the Schur basis state $\ket{\lambda,q_\lambda,p_\lambda}_{\rm
Sch}$ to the computational basis state
$\ket{\lambda,q_\lambda,p_\lambda}$ with $\lambda$, $q_\lambda$, and
$p_\lambda$ expressed as bit strings.  The dimensions of the irreps
${\bf p}_\lambda$ and ${\bf q}^d_\lambda$ vary with $\lambda$, so we
will need to pad the $|q_\lambda,p_\lambda\rangle$ registers when they
are expressed as bit strings.  We will label the padded basis as
$\ket{\lambda}\ket{q}\ket{p}$, explicitly dropping the $\lambda$
dependence.  Later in the paper we will show how to do this padding
efficiently with only a logarithmic spatial overhead.  We will refer to
the transform from the computational basis $|i_1,i_2,\dots,i_n\rangle$
to the basis of three strings $|\lambda\rangle|q\rangle|p\rangle$ as
the Schur transform. The Schur transform is shown schematically in
Fig.~\ref{fig:schur}.  Notice that just as the standard computational
basis $|i\rangle$ is arbitrary up to a unitary transform, the bases
for $\cQ_\lambda^d$ and $\cP_\lambda$ are also both arbitrary up to a
unitary transform, though we will later choose particular bases for
$\cQ_\lambda^d$ and $\cP_\lambda$.

{\em Example of the Schur transform---}Let $d=2$.  Then for $n=2$
there are two valid partitions, $\lambda_1=2,\lambda_2=0$ and
$\lambda_1=\lambda_2=1$.  Here the Schur transform corresponds to
the change of basis from the standard basis to the singlet and
triplet basis: $|\lambda=(1,1),q_\lambda=0,p_\lambda=0\rangle_{\rm
Sch}=\frac{1}{\sqrt{2}}(|01\rangle-|10\rangle)$,
$|\lambda=(2,0),q_\lambda=+1,p_\lambda=0\rangle_{\rm
Sch}=|00\rangle$,
$|\lambda=(2,0),q_\lambda=0,p_\lambda=0\rangle_{\rm Sch}=\frac{1}
{\sqrt{2}} (|01\rangle+|10\rangle)$, and
$|\lambda=(2,0),q_\lambda=-1,p_\lambda=0\rangle_{\rm
Sch}=|11\rangle$. Abstractly, then, the Schur transform then
corresponds to a transformation
\begin{equation}
{\bf U}_{\rm Sch}=\left.\begin{array}{c} |\lambda=(1,1),q_\lambda=0,p_\lambda=0\rangle_{\rm Sch} \\
|\lambda=(2,0),q_\lambda=+1,p_\lambda=0\rangle_{\rm Sch}
\\ |\lambda=(2,0),q_\lambda=0,p_\lambda=0\rangle_{\rm Sch}\\ |\lambda=(2,0),q_\lambda=-1,p_\lambda=0\rangle_{\rm
Sch}\end{array}\right \{ \overbrace{\left[\begin{array}{cccc} 0 & {1 \over
\sqrt{2}} & - {1 \over \sqrt{2}} & 0 \\ 1 & 0 & 0 & 0 \\ 0 &  {1 \over
\sqrt{2}} &  {1 \over \sqrt{2}} & 0 \\ 0 & 0 & 0 & 1
\end{array}\right]}^{|00\rangle~|01\rangle~|10\rangle~|11\rangle}
\end{equation}
It is easy to verify that the $\lambda=(1,1)$ subspace transforms as a
one dimensional irrep of ${\mathcal U}_2$ and as the alternating sign
irrep of ${\mathcal S}_2$ and that the $\lambda=(2,0)$ subspace
transforms as a three dimensional irrep of ${\mathcal U}_2$ and as the
trivial irrep of ${\mathcal S}_2$.  Notice that the labeling scheme
for the standard computational basis uses $2$ qubits while the
labeling scheme for the Schur basis uses more qubits (one such
labeling assigns one qubit to $|\lambda\rangle$, none to $|p\rangle$
and two qubits to $|q\rangle$).  Thus we see how padding will be
necessary to directly implement the Schur transform.

To see a more complicated example of the Schur basis, let $d=2$
and $n=3$.  There are again two valid partitions, $\lambda=(3,0)$
and $\lambda=(2,1)$.  The first of these partitions labels to
the trivial irrep of ${\mathcal S}_3$ and a $4$ dimensional irrep
of ${\mathcal U}_3$.  The corresponding Schur basis vectors can be
expressed as
\be\begin{split}
|\lambda=(3,0),q_\lambda=+3/2,p_\lambda=0\rangle_{\rm
Sch} & = |000\rangle \\
|\lambda=(3,0),q_\lambda=+1/2,p_\lambda=0\rangle_{\rm Sch}
&=
{1\over \sqrt{3}}\left(|001\rangle+|010\rangle+|100\rangle\right)\\
|\lambda=(3,0),q_\lambda=-1/2,p_\lambda=0\rangle_{\rm Sch}
&=
{1\over \sqrt{3}}\left(|011\rangle+|101\rangle+|110\rangle\right)\\
|\lambda=(3,0),q_\lambda=-3/2,p_\lambda=0\rangle_{\rm
Sch} &= |111\rangle.\end{split}\label{eq:3qubit-sym}\ee

The second of these partitions labels
a two dimensional irrep of ${\mathcal S}_3$ and a two dimensional
irrep of ${\mathcal U}_2$.  Its Schur basis states can be expressed
as
\be\begin{split}
|\lambda=(2,1),q_\lambda=+1/2,p_\lambda=0\rangle_{\rm Sch}&=
{1\over \sqrt{2}} \left(|100\rangle - |010\rangle\r)\\
|\lambda=(2,1),q_\lambda=-1/2,p_\lambda=0\rangle_{\rm Sch}&=
{1\over \sqrt{2}} \left(|101\rangle - |011\rangle\r)\\
|\lambda=(2,1),q_\lambda=+1/2,p_\lambda=1\rangle_{\rm Sch}&=
\sqrt{\frac{2}{3}}|001\> - \frac{|010\> + |100\>}{\sqrt{6}}\\
|\lambda=(2,1),q_\lambda=-1/2,p_\lambda=1\rangle_{\rm Sch}&=
\sqrt{\frac{2}{3}}|110\> - \frac{|101\> + |011\>}{\sqrt{6}}.
\end{split}\label{eq:3qubit-mixed}\ee
We can easily verify that \eqs{3qubit-sym}{3qubit-mixed} indeed
transform under $\cU_2$ and $\cS_3$ the way we expect; not so easy
however is generalizing this basis to any $n$ and $d$, let alone
coming up with a natural circuit relating this basis to the
computational basis.  However, note that $p_\lambda$ determines
whether the first two qubits are in a singlet or a triplet state.
This gives a hint of a recursive structure that we will exploit in
\sect{gz} to describe Schur bases for any choice of $n$ and $d$, and
in \sect{schur-circuit} to construct an efficient recursive algorithm for the
Schur transform.

\subsection{Constructing $\cQ_\lambda^d$ and $\cP_\lambda$ using Schur
duality}\label{sec:weyl-symmetrizer}
So far we have said little about the form of $\cQ_\lambda^d$
and $\cP_\lambda$, other than that they are indexed by partitions.  It
turns out that Schur duality gives a straightforward description of
the irreps of $\cU_d$ and $\cS_n$.  We will not use this explicit
description to construct the Schur transform, but it is
still helpful for understanding the irreps $\cQ_\lambda^d$
and $\cP_\lambda$.  As with the rest of this section, proofs and
further details can be found in \cite{Goodman:98a}.

We begin by expressing $\lambda\in\cI_{d,n}$ as a Young diagram in
which there are up to $d$ rows with $\lambda_i$ boxes in row $i$.  For
example, to the partition $(4,3,1,1)$ we associate the diagram
\be \yng(4,3,1,1). \label{eq:ferrers-example}\ee
Now we define a Young tableau $T$ of shape $\lambda$ to be a way of
filling the $n$ boxes of $\lambda$ with the integers $1,\ldots,n$,
using each number once and so that integers increase from left to
right and from top to bottom.  For example, one valid Young tableau
with shape $(4,3,1,1)$ is
$$\young(1467,258,3,9).$$
For any Young tableau $T$, define
$\operatorname{Row}(T)$ to be set of permutations obtained by
permuting the integers within each row of $T$; similarly define
$\operatorname{Col}(T)$ to be the permutations that leave each integer
in the same column of $T$.  Now we define the {\em Young symmetrizer}
$\Pi_{\lambda:T}$ to be an operator acting on $(\bbC^d)^{\ot n}$ as follows:
\be \Pi_{\lambda:T} := \frac{\dim\cP_\lambda}{n!}
\l(\sum_{c\in\operatorname{Col}(T)} \sgn(c)\bP(c)\r)
\l(\sum_{r\in\operatorname{Row}(T)} \bP(r)\r).\ee

It can be shown that the Young symmetrizer $\Pi_{\lambda:T}$ is a projection
operator whose support is a subspace isomorphic to $\cQ_\lambda^d$.
In particular $\Usch\Pi_{\lambda:T} \Usch^\dag = \oprod{\lambda} \ot
\oprod{y(T)} \ot \bI_{\cQ_\lambda^d}$ for some unit vector
$\ket{y(T)}\in\cP_\lambda$.  Moreover, these vectors $\ket{y(T)}$ form
a basis known as Young's natural basis, though the $\ket{y(T)}$ are
not orthogonal, so
we will usually not work with them in quantum circuits.

Using Young symmetrizers, we can now explore some more general
examples of $\cQ_\lambda^d$ and $\cP_\lambda$.  If $\lambda=(n)$, then
the only valid tableau is
$$ \young(12)\cdots\young(n).$$
The corresponding $\cS_n$-irrep $\cP_{(n)}$ is trivial and the
$\cU_d$-irrep is given by the action of $\bQ$ on the totally symmetric
subspace of $(\bbC^d)^{\ot n}$, i.e. $\{\ket{v} : \bP(s)\ket{v}=\ket{v}
\forall s\in\cS_n\}$.  On the other hand, if $\lambda=(1^n)$, meaning
$(1,1,\ldots, 1)$ ($n$ times), then the only valid tableau is
$$ \begin{array}{c}\young(1,2)\\\vdots\\\young(n)\end{array}.$$
The $\cS_n$-irrep $\cP_{(1^n)}$ is still one-dimensional, but now
corresponds to the sign irrep of $\cS_n$, mapping $s$ to $\sgn(s)$.
The $\cU_d$-irrep $\cQ_{(1^n)}^d$ is equivalent to the totally
antisymmetric subspace of $(\bbC^d)^{\ot n}$, i.e. $\{\ket{v} :
\bP(s)\ket{v}=\sgn(s)\ket{v} \forall s\in\cS_n\}$.  Note that if
$d>n$, then this subspace is zero-dimensional, corresponding to the
restriction that irreps of $\cU_d$ are indexed only by partitions with
$\leq d$ rows.

Other explicit examples of $\cU_d$ and $\cS_n$ irreps are presented from a
particle physics perspective in \cite{Georgi99}.  We also give more
examples in \sect{gz-yy}, when we introduce explicit bases for
$\cQ_\lambda^d$ and $\cP_\lambda$.

\subsection{Applications of the Schur Transform}
\label{sec:schur-qit-apps}

The Schur transform is useful in a surprisingly large number of
quantum information protocols.  Here we, review these
applications, with particular attention to the use of the Schur
transform circuit in each protocol.  We emphasize again that our
construction of the Schur transform simultaneously makes all of
these tasks computationally efficient.

\subsubsection{Spectrum and state estimation}

Suppose we are given many copies of an unknown mixed quantum
state, $\bmath{\rho}^{\otimes n}$.  An important task is to obtain
an estimate for the spectrum of $\bmath{\rho}$ from these $n$
copies.  An asymptotically good estimate (in the sense of large
deviation rate) for the spectrum of $\bmath{\rho}$ can be obtained
by applying the Schur transform, measuring $\lambda$ and taking
the spectrum estimate to be $(\lambda_1/n,\ldots,
\lambda_d/n)$\cite{Keyl:01a,Vidal:99a}.  Thus an efficient
implementation of the Schur transform will efficiently implement
the spectrum estimating protocol (note that it is efficient in
$d$, not in $\log(d)$).  Estimating $\bmath{\rho}$ reduces to
measuring $|\lambda\rangle$ and $|q\rangle$, but optimal
estimators have only been explicitly constructed for the case of
$d=2$\cite{Gill:02a}.  Further, optimal quantum hypothesis testing
can be obtained by a similar protocol\cite{Hayashi:02d}.

\subsubsection{Universal distortion-free entanglement concentration}

Let $|\psi\rangle_{AB}$ be a bipartite partially entangled state
shared between two parties, $A$ and $B$. Suppose we are given many
copies of $|\psi\rangle_{AB}$ and we want to transform these
states into copies of a maximally entangled state using only local
operations and classical communication.  Further, suppose that we
wish this protocol to work when neither $A$ nor $B$ know the state
$|\psi \rangle_{AB}$.  Such a scheme is called a universal
(meaning it works with unknown states $|\psi\rangle_{AB}$)
entanglement concentration protocol, as opposed to the original
entanglement concentration protocol described by Bennett {\em
et.al.}\cite{Bennett:96b}.  Further we also would like the
scheme to produce perfect maximally entangled states, i.e. to be
distortion free.  Universal distortion-free entanglement
concentration can be performed\cite{Hayashi:02a} by both parties
performing Schur transforms on their $n$ halves of
$|\psi\rangle_{AB}$, measuring their $|\lambda\rangle$, discarding
$|q\rangle$ and retaining $|p\rangle$.  The two parties will now
share a maximally entangled state of varying dimension depending
on what $\lambda$ was measured.  This dimension asymptotes to
$2^{n H}$, where $H$ is the entropy of one of the parties' reduced
mixed states.

\subsubsection{Universal Compression with Optimal Overflow
Exponent}

Measuring $|\lambda\rangle$ weakly so as to cause little disturbance, together
with appropriate relabeling, comprises a universal compression algorithm with
optimal overflow exponent (rate of decrease of the probability that the
algorithm will output a state that is much too
large)\cite{Hayashi:02b,Hayashi:02c}.

\subsubsection{Encoding and decoding into decoherence-free subsystems}

Further applications of the Schur transform include encoding into
decoherence-free
subsystems\cite{Zanardi:97a,Knill:00a,Kempe:01a,Bacon:01a}.
Decoherence-free subsystems are subspaces of a system's Hilbert
space which are immune to decoherence due to a symmetry of the
system-environment interaction.  For the case where the
environment couples identically to all systems, information can be
protected from decoherence by encoding into the $|p_\lambda
\rangle$ basis.  We can use the inverse Schur transform (which, as
a circuit can be implemented by reversing the order of all gate
elements and replacing them with their inverses) to perform this
encoding: simply feed in
the appropriate $|\lambda\rangle$ with the state to be encoded
into the $|p\rangle$ register and any state into the $|q\rangle$
register into the inverse Schur transform.  Decoding can similarly
be performed using the Schur transform.

\subsubsection{Communication without a shared reference frame}

An application of the concepts of decoherence-free subsystems
comes about when two parties wish to communicate (in either a
classical or quantum manner) when the parties do not share a
reference frame.  The effect of not sharing a reference frame is
the same as the effect of collective decoherence (the same random
unitary rotation has been applied to each subsystem).  Thus
encoding information into the $|p\rangle$ register will allow this
information to be communicated in spite of the fact that the two
parties do not share a reference frame\cite{Bartlett:03a}.  Just
as with decoherence-free subsystems, this encoding and
decoding can be done with the Schur transform.

\section{Subgroup adapted bases and the Schur basis} \label{sec:gz}

In the last section, we defined the Schur transform in a way that left
the basis almost completely arbitrary.  To construct a quantum circuit for
the Schur transform, we will need to explicitly specify the Schur
basis.  Since we want the Schur basis to be of the form
$\ket{\lambda,q,p}$, our task reduces to specifying orthonormal
bases for $\cQ_\lambda^d$ and $\cP_\lambda$.   We will call these bases
$Q_\lambda^d$ and $P_\lambda$, respectively.

We will choose $Q_\lambda^d$ and $P_\lambda$ to both be a type of
basis known as a {\em subgroup-adapted} basis.  In \sect{subgp-adapted}
we describe the general theory of subgroup-adapted bases, and in
\sect{gz-yy}, we will describe subgroup-adapted bases for
$\cQ_\lambda^d$ and $\cP_\lambda$.  As we will later see, the properties of
these bases are intimately related to the structure of the algorithms
that work with them.  In this section, we will show how the bases
can be stored on a quantum computer with a small amount of padding,
and in the following sections we will show how the subgroup-adapted
bases described here enable efficient implementations of
Clebsch-Gordan and Schur duality transforms.

\subsection{Subgroup Adapted Bases}\label{sec:subgp-adapted}

Here we review the basic idea of a subgroup adapted basis.  We
assume that all groups we talk about are finite or compact Lie groups.
Suppose $(\br,V)$ is an irrep of a group $\cG$ and $\cH$ is a proper
subgroup of $\cG$.  We will construct a basis for $V$ via the
representations of $\cH$.

Begin by restricting the input of $\br$ to $\cH$ to obtain a
representation of $\cH$, which we call $(\br|_\cH,V\da_\cH)$.  Note
that unlike $V$, $V\da_\cH$ may be reducible.  In fact, if we let
$(\br'_\alpha,V'_\alpha)$ denote the irreps of $\cH$, then $V\da_\cH$
will decompose under the action of $\cH$ as
\be V\da_\cH \stackrel{\cH}{\cong} \bigoplus_{\alpha\in\hat{\cH}}
\bbC^{n_\alpha} \ot V'_\alpha   \ee 
or equivalently, $\br|_\cH$ decomposes as
\begin{equation}
\br(h) = {\bf r}\da_{\mathcal H}(h) \cong \bigoplus_{\alpha\in\hat{\cH}}
 {\bf I}_{n_\alpha} \otimes
{\bf r}'_\alpha(h) \label{eq:restrict}
\end{equation}
where $\hat{\cH}$ runs over a complete set of inequivalent irreps of
$\cH$ and $n_\alpha$ is the {\em branching multiplicity} of the irrep
labeled by $\alpha$.  Note that since $\br$ is a unitary
representation, the subspaces corresponding to different irreps of
$\cH$ are orthogonal.  Thus, the problem of finding an orthonormal
basis for $V$ now reduces to
the problem of (1) finding an orthonormal basis for each irrep of $\cH$,
$V_\alpha'$ and (2) finding orthonormal bases for the multiplicity spaces
$\bbC^{n_\alpha}$.  The case when all the $n_\alpha$ are either 0 or 1
is known as {\em multiplicity-free branching}.  When this occurs, we
only need to determine which irreps occur in the decomposition of $V$,
and find bases for them.

Now consider a group ${\mathcal G}$ along with a tower of subgroups
${\mathcal G}={\mathcal G}_1\supset {\mathcal G}_2 \supset \dots
\supset {\mathcal G}_{k-1} \supset {\mathcal G}_{k}=\{e\}$ where $\{e\}$
is the trivial subgroup consisting of only the identity element.  For
each $\cG_i$, denote its irreps by $V_\alpha^i$, for
$\alpha\in\hat{\cG}_i$.   Any irrep $V_{\alpha_1}^1$ of
$\cG={\mathcal G}_1$ decomposes under restriction to $\cG_2$ into
$\cG_2$-irreps: say that $V_{\alpha_2}^2$ appears
$n_{\alpha_1,\alpha_2}$ times.  We can then look at these irreps of
${\mathcal G}_2$, consider their restriction to $\cG_3$ and decompose
them into different irreps of ${\mathcal G}_3$. Carrying on in such a
manner down this tower of subgroups will yield a labeling for
subspaces corresponding to each of these restrictions.  Moreover, if
we choose orthonormal bases for the multiplicity spaces, this will
induce an orthonormal basis for $\cG$. 
This basis is known as a {\em subgroup-adapted basis} and basis
vectors have the form
$\ket{\alpha_2,m_2,\alpha_3,m_3,\ldots,\alpha_n,m_n}$, where
$\ket{m_i}$ is a basis vector for the
($n_{\alpha_{i-1},\alpha_i}$-dimensional) multiplicity space of
$V_{\alpha_i}^i$ in $V_{\alpha_{i-1}}^{i-1}$.

If the branching for each $\cG_{i+1}\subset \cG_i$ is
multiplicity-free, then we say that the tower of subgroups is {\em
canonical}.  In this case, the subgroup adapted basis takes the
particularly simple form of $\ket{\alpha_2,\ldots,\alpha_n}$, where
each $\alpha_i\in\hat{\cG}_i$ and $\alpha_{i+1}$ appears in the
decomposition of $V_{\alpha_i}\da_{\cG_{i+1}}$.  Often we include the
original irrep label $\alpha=\alpha_1$ as well:
$|\alpha_1,\alpha_2,\dots,\alpha_k\rangle$.  This means that there
exists a basis whose vectors are completely determined (up to an
arbitrary choice of phase) by which irreps of $G_1,\ldots,G_k$ they
transform according to.  Notice that a basis for the irrep $V_\alpha$
does not consist of all possible irrep labels $\alpha_i$, but instead
only those which can appear under the restriction which defines the
basis.

The simple recursive structure of subgroup adapted bases makes them
well-suited to performing explicit computations.  Thus, for example,
subgroup adapted bases play a major role in efficient quantum circuits
for the Fourier transform over many nonabelian groups\cite{Moore:03a}.

\subsection{Explicit orthonormal bases for $\cQ_\lambda^d$ and
$\cP_\lambda$} \label{sec:gz-yy}

In this section we describe canonical towers of subgroups for $\cU_d$
and $\cS_n$, which give rise to subgroup-adapted bases for the irreps
$\cQ_\lambda^d$ and $\cP_\lambda$.  These bases go by many names: for
$\cU_d$ (and other Lie groups) the basis is called the Gel'fand-Zetlin
basis (following \cite{Gelfand:50a}) and we denote it by $Q_\lambda^d$,
while for $\cS_n$ it is called the Young-Yamanouchi basis, or
sometimes Young's orthogonal basis (see
\cite{James:81a} for a good review of its properties) and is denoted
$P_\lambda$.  The
constructions and corresponding branching rules are quite simple, but
for proofs we again refer the reader to \cite{Goodman:98a}.

{\em The Gel'fand-Zetlin basis for $\cQ_\lambda^d$---}
For $\cU_d$, it turns out that the chain of subgroups $\{1\}=\cU_0\subset
\cU_1 \subset \ldots \subset \cU_{d-1} \subset \cU_d$ is a canonical
tower.  For $c<d$, the subgroup $\cU_c$ is embedded in $\cU_d$ by
$\cU_c := \{u\in\cU_d : u|i\> = |i\> \text{ for } i = c+1,\ldots,
d\}$.  In other words, it corresponds to matrices of the form
\be U\oplus I_{d-c} :=
\l(\begin{array}{c|c} u  & 0\\ \hline
\\ 0 & I_{d-c}\end{array}\r)
,\ee where $u$ is a $c\times c$ unitary matrix.

Since the branching from $\cU_d$ to $\cU_{d-1}$ is multiplicity-free,
we obtain a subgroup-adapted basis $Q_\lambda^d$, which is known as
the {\em Gel'fand-Zetlin} (GZ) basis.  Our only free choice in a GZ
basis is the initial choice of basis $\ket{1},\ldots,\ket{d}$ for
$\bbC^d$ which determines the canonical tower of subgroups
$\cU_1\subset \ldots \subset \cU_d$.  Once we have chosen this basis,
specifying $Q_\lambda^d$ reduces to knowing which irreps
$\cQ_\mu^{d-1}$ appear in the decomposition of
$\cQ_\lambda^d\da_{\cU_{d-1}}$.  Recall that the irreps of $\cU_d$ are
labeled by elements of $\cI_{d,n}$ with $n$ arbitrary.  This set can
be denoted by $\bbZ_{++}^d := \cup_n
\cI_{d,n} = \{\lambda\in\bbZ^d :
\lambda_1 \geq \ldots \geq \lambda_d \geq 0\}$.  For
$\mu\in\bbZ_{++}^{d-1}, \lambda\in\bbZ_{++}^d$, we say that $\mu$ {\em
interlaces} $\lambda$ and write $\mu\interlaces\lambda$ whenever $\lambda_1
\geq \mu_1 \geq \lambda_2 \ldots \geq \lambda_{d-1} \geq \mu_{d-1}
\geq \lambda_d$.  In terms of Young diagrams, this means that $\mu$
is a valid partition (i.e. a nonnegative, nonincreasing sequence)
obtained from removing zero or one boxes from each column of
$\lambda$.  For example, if $\lambda=(4,3,1,1)$ (as in
\eq{ferrers-example}), then $\mu\interlaces\lambda$ can be obtained by
removing any subset of the marked boxes below, although if the box
marked $*$ on the second line is removed, then the other marked box on
that line must also be removed.
\be \young(\hfil\hfil\hfil\times,\hfil *\times,\hfil,\times)\ee

Thus a basis vector in $Q_\lambda^d$ corresponds to a sequence of
partitions $q=(q_d\!=\!\lambda, \ldots, q_1)$ such that $q_1\interlaces q_2
\interlaces \ldots \interlaces q_d$ and $q_j\in\bbZ^j_{++}$ for $j=1,\ldots,d$.
Again using $\lambda=(4,3,1,1)$ as an example, and choosing $d=5$ (any
$d\geq 4$ is possible), we might have the sequence
\be \begin{array}{cccccccccc}
\Yvcentermath1
\yng(4,3,1,1)& \reverselaces& \Yvcentermath1\yng(3,3,1) &\reverselaces &
\Yvcentermath1\yng(3,3,1)&
\reverselaces& \Yvcentermath1\yng(3,1)& \reverselaces & \Yvcentermath1\yng(2)\\
q_5&&q_4&&q_3&&q_2&&q_1
\end{array} \label{eq:ex-partition-chain}\ee
Observe that it is possible in some steps not to remove any boxes, as
long as $q_j$ has no more than $j$ rows.

In order to work with the Gel'fand-Zetlin basis vectors on a quantum
computer, we will need an efficient method to write them down.
Typically, we think of $d$ as constant and express our resource use in
terms of $n$.  Then an element of $\cI_{d,n}$ can be expressed with
$d\log(n+1)$ bits, since it consists of $d$ integers between $0$ and
$n$.  (This is a crude upper bound on $|\cI_{d,n}|=\binom{n+d-1}{d-1}$,
but for constant $d$ it is good enough for our purposes.)  A
Gel'fand-Zetlin basis vector then requires no more than $d^2\log(n+1)$
bits, since it can be expressed as $d$ partitions of integers no
greater than $n$ into $\leq d$ parts.  Here, we assume that all
partitions have arisen from a decomposition of $(\bbC^d)^{\ot n}$, so
that no Young diagram has more than $n$ boxes.  Unless otherwise
specified, our algorithms will use this encoding of the GZ basis
vectors.

It is also possible to express GZ basis vectors in a more visually
appealing way by writing numbers in the boxes of a Young diagram.  If
$q_1\interlaces \ldots \interlaces q_d$ is a chain of partitions, then
we write the number $j$ in each box contained in $q_j$ but not
$q_{j-1}$ (with $q_0=(0)$).  For example, the sequence in
\eq{ex-partition-chain} would be denoted
\be \young(1125,233,3,5).
\label{eq:ssyt-example}\ee
Equivalently, any method of filling a Young diagram with numbers from
$1,\ldots,d$ corresponds to a valid chain of irreps as long as the
numbers are nondecreasing from left to right and are strictly
increasing from top to bottom.  The resulting diagram is known as a
{\em semi-standard Young tableau} and gives another way of encoding a
GZ basis vector; this time using $n\log d$ bits.  (It turns out the
actual dimension of $\cQ_\lambda^d$ is $\l[\prod_{1\leq i<j\leq d}
(\lambda_i - \lambda_j + j - i)\r]/\l[\prod_{m=1}^d m!\r]$, and later
in this section we will give an algorithm for
efficiently encoding a GZ basis vector in the optimal $\lceil \log
\dim\cQ_\lambda^d\rceil$ qubits.  However, this is not necessary for
most applications.)

{\em Example: irreps of $\cU_2$---}  To ground the above discussion in
an example more familiar to physicists, we show how the GZ basis for
$\cU_2$ irreps corresponds to states of definite angular momentum
along one axis.  An irrep of $\cU_2$ is labeled by two integers
$(\lambda_1,\lambda_2)$ such that $\lambda_1+\lambda_2=n$ and
$\lambda_1 \geq \lambda_2 \geq 0$.  A GZ basis vector for
$\cQ_\lambda^2$ has $\lambda_2+m$ 1's in the first row, followed by
$\lambda_1-(\lambda_2+m)$ 2's in the first row and $\lambda_2$ 2's in
the second row, where $m$ ranges from 0 to $\lambda_1-\lambda_2$.
  This arrangement is necessary to satisfy
the constraint that numbers are strictly increasing from top to bottom
and are nondecreasing 
from left to right.  Since the GZ basis vectors are completely specified
by $m$, we can label the vector $\ket{(\lambda_1,\lambda_2) ;
(\lambda_2 + m)}\in 
Q_\lambda^2$ simply by $\ket{m}$.  For example, $\lambda=(9,4)$ and $m=2$ would
look like
\be \young(111111222,2222).\ee

Now observe that $\dim\cQ_\lambda^2 = \lambda_1 - \lambda_2 +1$, a
fact which is consistent with having angular momentum
$J=(\lambda_1-\lambda_2)/2$.  We claim that $m$ corresponds to the $Z$
component of angular momentum (specifically, the $Z$ component of
angular momentum is $m-J = m - (\lambda_1-\lambda_2)/2$).  To see
this, first note that $\cU_1$ acts 
on a GZ basis vector $\ket{m}$ according to the representation $x \ra
x^{\lambda_2 + m}$, for $x\in\cU_1$; equivalently
$\bq_\lambda^2\l(\l(\begin{smallmatrix}x&0\\0&1\end{smallmatrix}\r)\r)\ket{m}
= x^{\lambda_2+m}\ket{m}$. Since $\bq_\lambda^2(yI_2)\ket{m} =
y^n\ket{m} = y^{\lambda_1+\lambda_2}\ket{m}$, we can find the action
of $e^{i\theta\sigma_z} = \l(\begin{smallmatrix}e^{2i\theta}&0\\0 &
1\end{smallmatrix} \r)
\l(\begin{smallmatrix}e^{-i\theta} & 0 \\ 0 &
e^{-i\theta}\end{smallmatrix}\r)$ on 
$\ket{m}$.  We do this by combining the above arguments to find that
$\bq_\lambda^2(e^{i\theta\sigma_z})\ket{m} =
e^{2i\theta(\lambda_2+m)}e^{-i\theta(\lambda_1+\lambda_2)}\ket{m} =
e^{2i\theta(m-J)}\ket{m}$.  Thus we obtain the desired action of
a $Z$ rotation on a particle with total angular momentum $J$ and
$Z$-component of angular momentum $m$.

{\em Example: The defining irrep of $\cU_d$---}  The simplest
nontrivial irrep of $\cU_d$ is its action on $\bbC^d$.  This
corresponds to the partition $(1)$, so we say that
$(\bq_{(1)}^d,\cQ_{(1)}^d)$ is the {\em defining irrep} of $\cU_d$
with $\cQ_{(1)}^d = \bbC^d$ and $\bq_{(1)}^d(U) = U$.
Let $\ket{1},\ldots,\ket{d}$ be an orthonormal basis for $\bbC^d$
corresponding to the canonical tower of subgroups $\cU_1\subset \cdots
\subset \cU_d$.  It turns out that this is already a GZ basis.  To see
this, note
that $\cQ_{(1)}^d\da_{\cU_{d-1}}\stackrel{\cU_{d-1}}{\cong}
\cQ_{(0)}^{d-1}
\oplus \cQ_{(1)}^{d-1}$.  This is because $\ket{d}$ generates
$\cQ_{(0)}^{d-1}$, a trivial irrep of $\cU_{d-1}$; and
$\ket{1},\ldots,\ket{d-1}$ generate $\cQ_{(1)}^{d-1}$, a defining
irrep of $\cU_{d-1}$.  Another way to say this is that $\ket{j}$ is
acted on according to the trivial irrep of $\cU_1,\ldots,\cU_{j-1}$
and according to the defining irrep of $\cU_j,\ldots,\cU_d$.  Thus
$\ket{j}$ corresponds to the chain of partitions
$\{(0)^{j-1},(1)^{d-j+1}\}$.  We will return to this example many
times in the rest of the paper.

{\em The Young-Yamanouchi basis for $\cP_\lambda$---}
The situation for $\cS_n$ is quite similar.  Our chain of subgroups is
$\{e\} = \cS_1 \subset \cS_2 \subset \ldots \subset \cS_n$, where for
$m<n$ we define $\cS_m\subset \cS_n$ to be the permutations in $\cS_n$
which leave the last $n-m$ elements fixed.  For example, if $n=3$,
then ${\mathcal S}_3=\{e,(12),(23),(13),(123),(321)\}$, ${\mathcal
S}_2=\{e,(12)\}$, and ${\mathcal S}_1=\{e\}$.  Recall that the
irreps of ${\mathcal S}_n$ can be labeled by $\cI_n=\cI_{n,n}$: the
partitions of $n$ into $\leq n$ parts.

Again, the branching from $\cS_n$ to $\cS_{n-1}$
is multiplicity-free, so to determine an orthonormal basis $P_\lambda$
for the space $\cP_\lambda$ we need only know which irreps occur in the
decomposition of $\cP_\lambda\da_{\cS_{n-1}}$.  It turns out that the
branching rule is given by finding all ways to remove one box from
$\lambda$ while leaving a valid partition.  Denote the set of such
partitions $\lambda-\Box$.  Formally, $\lambda-\Box := \cI_n \cap
\{\lambda-e_j : j=1,\ldots,n\}$, where $e_j$ is the unit vector in
$\bbZ^n$ with a one in the $j^{\text{th}}$ position and zeroes
elsewhere.  Thus, the general branching rule is
\be \cP_\lambda\da_{\cS_{n-1}} \iso{\cS_{n-1}}
\bigoplus_{\mu\in\lambda-\Box} \cP_\mu.
\label{eq:cP-branching}\ee
For example, if $\lambda=(3,2,1)$, we might have the chain
of partitions:
\be \begin{array}{cccccccccccc}
\Yvcentermath1\yng(3,2,1) & \ra & \Yvcentermath1\yng(2,2,1) & \ra &
\Yvcentermath1\yng(2,2) & \ra & \Yvcentermath1\yng(2,1) & \ra &
\Yvcentermath1\yng(1,1) & \ra & \Yvcentermath1\yng(1)\\
n=6 && n=5 && n=4 && n=3 && n=2 && n=1\end{array}\ee
Again, we can concisely label this chain by writing the number $j$ in
the box that is removed when restricting from $\cS_j$ to $\cS_{j-1}$.
The above example would then be
\be \young(136,24,5).\ee
Note that the valid methods of
filling a Young diagram are slightly different than for the $\cU_d$
case.  Now we use each integer in $1,\ldots,n$ exactly once such that
the numbers are increasing from left to right and from top to bottom.
The resulting tableau is called a {\em standard Young tableau}.
(The same filling scheme appeared in the description of Young's
natural representation in \sect{weyl-symmetrizer}, but the resulting
basis states are of course quite different.)

This gives rise to a straightforward, but inefficient, method of
writing an element
of $P_\lambda$ using $\log n!$ bits.  However, for applications such as
data compression\cite{Hayashi:02b,Hayashi:02c} we will need an
encoding which gives us closer to the optimal $\log P_\lambda$
bits.  First we note an exact (and efficiently computable) expression
for $|P_\lambda|=\dim\cP_\lambda$:
\be \dim \cP_\lambda =
\frac{n!}{\lambda_1+d-1!\lambda_2+d-2!\cdots \lambda_d!}
\prod_{1\leq i<j\leq d} (\lambda_i - \lambda_j + j - i).
\label{eq:cP-dim}\ee
Now we would like to efficiently and
reversibly map an element of $P_\lambda$ (thought of as a chain of
partitions $p=(p_n=\lambda,\ldots,p_1=(1))\in P_\lambda$, with $p_j
\in p_{j+1}-\Box$) to an integer in $[|P_\lambda|] :=
\{1,\ldots,|P_\lambda|\}$.  We will construct
this bijection $f_n:P_\lambda \ra [|P_\lambda|]$ by defining an ordering on
$P_\lambda$ and setting $f_n(p):=|\{p'\in P_\lambda : p'\leq p\}|$.
First fix an arbitrary, but easily computable, 
(total) ordering on partitions in $\cI_n$ for each $n$; for example,
lexicographical order.  This
induces an ordering on $P_\lambda$ if we rank a basis vector $p\in
P_\lambda$ first according to $p_{n-1}$, using the order on partitions
we have chosen, then according to $p_{n-2}$ and so on.  We skip $p_n$,
since it is always equal to $\lambda$.  In other words, for $p,p'\in
P_\lambda$, $p>p'$ if $p_{n-1}>p_{n-1}'$ {\em or} $p_{n-1}=p_{n-1}'$
and $p_{n-2}>p_{n-2}'$ {\em or} $p_{n-1}=p_{n-1}'$,
$p_{n-2}=p_{n-2}'$ and $p_{n-3}>p_{n-3}'$, and so on.
Thus $f_n:P_\lambda \ra [|P_\lambda|]$ can be easily verified to be
\be f_n(p) = f_n(p_1,\ldots,p_n) :=
1 + \sum_{k=2}^n\sum_{\substack{\mu\in p_k-\Box\\\mu<p_{k-1}}}
\dim \cP_\mu.
\label{eq:Plambda-relabel}\ee 
Thus $f_n$ is an injective map from $P_\lambda$
to $[|P_\lambda|]$.
Moreover, since there are $O(n^2)$ terms in \eq{Plambda-relabel} and
\eq{cP-dim} gives an efficient way to calculate each
$|P_\lambda|$, this mapping can be performed in time polynomial in $n$.


\section{The Clebsch-Gordan transform and efficient circuits for the
Schur transform}\label{sec:schur-circuit}

In this section, we describe an efficient circuit for the Schur
transform $\Usch$.   To do so, we first describe the Clebsch-Gordan
transform, which decomposes a Kronecker product of $\cU_d$-irreps
$\cQ_\mu^d \ot \cQ_\nu^d$ into a direct sum of other $\cU_d$-irreps.
We defer the algorithm for the Clebsch-Gordan transform to
\sect{cg-construct}, 
but give a description of its properties in \sect{cg-def}.  Then in
\sect{schurcg}, we show how to construct the Schur transform by cascading
a series of Clebsch-Gordan transforms and performing reversible
classical manipulations of $P_\lambda$ and $Q_\lambda^d$.

\subsection{The Clebsch-Gordan Series and Transform}\label{sec:cg-def}

Suppose we have two irreps of $\cU_d$ given by the partitions $\mu$ and $\nu$:
$\cQ_\mu^d$ and $\cQ_\nu^d$.  The tensor product of these irreps $\cQ_\mu^d \ot
\cQ_\nu^d$ (with representation matrices $\bq_\mu^d(U) \ot \bq_\nu^d(U)$ for
$U\in\cU_d$) is a new representation of $\cU_d$.  This new representation will
generally be reducible into irreps of $\cU_d$; following \eq{hom-decomp} we
have
\begin{equation}
\cQ_\mu^d \ot \cQ_\nu^d \cong
\bigoplus_{\lambda\in\bbZ_{++}^d} \cQ_\lambda^d \ot
\Hom(\cQ_\lambda^d, \cQ_\mu^d \ot \cQ_\nu^d)^{\cU_d}
\label{eq:cg-decomp}
\end{equation}
This decomposition is referred to as the Clebsch-Gordan series.
Setting $N_{\mu\nu}^\lambda = \dim \Hom(\cQ_\lambda^d, \cQ_\mu^d \ot
\cQ_\nu^d)^{\cU_d}$, we obtain the corresponding
decomposition of the representation matrices as
\begin{equation}
\bq_\mu^d(U) \ot \bq_\nu^d(U) \cong
\bigoplus_{\lambda\in\bbZ_{++}^d} \bq_\lambda^d(U) \ot
I_{N_{\mu\nu}^\lambda}
\label{eq:cg-matrix}\end{equation}
The unitary matrix which maps the LHS of \eq{cg-matrix} to the RHS is
known as the Clebsch-Gordan transform and we denote it
$\Ucg^{\mu,\nu}$.  It maps vectors of the form
$\ket{q_\mu}\ket{q_\nu}\in \cQ_{\mu}^d\ot\cQ_\nu^d$ to superpositions
of vectors of the form $\ket{\lambda}\ket{q_\lambda}\ket{\alpha}$,
where $\lambda\in\bbZ_{++}^d$, $\ket{q_\lambda}\in \cQ_\lambda^d$ and
$\alpha \in \Hom(\cQ_\lambda^d, \cQ_\mu^d \ot \cQ_\nu^d)^{\cU_d}$.

The multiplicity space $\Hom(\cQ_\lambda^d, \cQ_\mu^d \ot
\cQ_\nu^d)^{\cU_d}$ plays a crucial role in the CG transform.  In
particular, the inverse CG transform $(\Ucg^{\mu,\nu})^\dag$ is given
simply by
\be (\Ucg^{\mu,\nu})^\dag \ket{q_\lambda}\ket{\alpha}
 = \alpha \ket{q_\lambda}.
\label{eq:Ucg-dag}\ee
Note that on the LHS, we interpret $\ket{\alpha}$ as a vector in the
multiplicity space $\Hom(\cQ_\lambda^d, \cQ_\mu^d \ot
\cQ_\nu^d)^{\cU_d}$, and on the RHS we treat $\alpha$ as an operator.
These are normalized such that $\ket{\alpha}$ is a unit vector if and
only if $\alpha$ is an isometry.

We now specialize to the case of tensoring in the defining irrep
$\cQ_{(1)}^d$, for
which the CG transform is particularly simple.  Recall that for
$\lambda\in\bbZ_{++}^d$ and $1\leq j\leq d$ we have $\lambda+e_j =
(\lambda_1, \ldots, \lambda_{j-1}, \lambda_j+1, \lambda_{j+1}, \ldots,
\lambda_d)$.   This is not always a
valid partition, i.e. if $\lambda'=\lambda+e_j$, the condition
$\lambda_{j-1}' \geq \lambda_{j}'$ might not
hold.  Recall that the valid partitions (of any integer) are given by
the set $\bbZ_{++}^d$.  Then the Clebsch-Gordan series we are
interested in is given by
\begin{equation}
\cQ_{\lambda}^d \ot \cQ_{(1)}^d \stackrel{\cU_d}{\cong}
\bigoplus_{\substack{j=1,\ldots d\\ \lambda+e_j\in\bbZ_{++}^d}}
\cQ_{\lambda+e_j}^d
\label{eq:cg-add-one}\end{equation}
This is the ``add a single box'' prescription for tensoring in a defining
representation of $\cU_d$: we add a single box to a Young diagram and if the new Young diagram is a valid Young diagram (i.e. corresponds to a
valid partition), then this irrep appears in the Clebsch-Gordan series.
For example if $\lambda=(3,2,1)$ then
\begin{equation}
\cQ_{(3,2,1)}^3 \otimes \cQ_{(1)}^3 \stackrel{\cU_3}{\cong}
\cQ_{(4,2,1)}^3 \oplus \cQ_{(3,3,1)}^3 \oplus \cQ_{(3,2,2)}^3
\end{equation}
or in Young diagram form
\begin{equation}
\Yvcentermath1 \yng(3,2,1) \otimes \yng(1)
\stackrel{\cU_3}{\cong}
 \yng(4,2,1) \oplus \yng(3,3,1)\oplus \yng(3,2,2)
\end{equation}
Note that if we had $d>3$, then the partition $(3,2,1,1)$ would also
appear.

We now seek to define the CG transform as a quantum circuit.  We
specialize to the case where one of the input irreps is the defining
irrep, but allow the other irrep to be specified by a quantum input.
The resulting CG  transform is defined as:
\be \Ucg = \sum_{\lambda\in\bbZ_{++}^d} \oprod{\lambda}
\ot \Ucg^{\lambda,(1)}.\ee
This takes as input a state of the form $\ket{\lambda}\ket{q}\ket{i}$,
for $\lambda\in\bbZ_{++}^d$, $\ket{q}\in Q_\lambda^d$ and $i\in [d]$.
The output is a superposition over vectors
$\ket{\lambda}\ket{\lambda'}\ket{q'}$, where $\lambda'=\lambda+e_j\in
\bbZ_{++}^d$, $j\in [d]$ and $\ket{q'}\in Q_{\lambda'}^d$.
Equivalently, we could output $\ket{\lambda}\ket{j}\ket{q'}$ or
$\ket{j}\ket{\lambda'}\ket{q'}$, since $(\lambda,\lambda')$,
$(\lambda,j)$ and $(\lambda',j)$ are all trivially related via
reversible classical circuits.

To better understand the input space of $\Ucg$, we introduce the {\em
model representation} $\cQ_*^d :=
\bigoplus_{\lambda\in\bbZ_{++}^d} \cQ_\lambda^d$, with corresponding
matrix $\bq_*^d(U) = \sum_\lambda \oprod{\lambda}\ot
\bq_\lambda^d(U)$.  The model representation (also sometimes called
the {\em Schwinger representation}) is infinite dimensional and
contains each irrep once.\footnote{By contrast, $L^2(\cU_d)$, which
will we not use, contains $\cQ_\lambda^d$ with multiplicity
$\dim\cQ_\lambda^d$} Its basis vectors are of the form
$\ket{\lambda,q}$ for $\lambda\in\bbZ_{++}^d$ and $\ket{q}\in
Q_\lambda^d$.  Since $\cQ_*^d$ is infinite-dimensional, we cannot
store it on a quantum computer and in this paper work only with
representations $\cQ_\lambda^d$ with $|\lambda|\leq n$; nevertheless
$\cQ_*^d$ is a useful abstraction.

Thus $\Ucg$ decomposes $\cQ_*^d \ot \cQ_{(1)}^d$ into irreps.  There
are two important things to notice about this version of the CG
transform.  First is that it operates simultaneously on different
input irreps.  Second is that different input irreps must remain
orthogonal, so in order to to maintain unitarity $\Ucg$ needs to keep
the information of which irrep we started with.  However, since
$\lambda' = \lambda + e_j$, this information requires only storing
some $j\in [d]$.  Thus, $\Ucg$ is a map from $\cQ_*^d \ot \bbC^d$ to
$\cQ_*^d \ot \bbC^d$, where the $\bbC^d$ in the input is the defining
representation and the $\bbC^d$ in the output tracks which irrep we
started with.

\begin{figure}[ht]
\begin{centering}
\leavevmode\xymatrix{
|\lambda\> & \gnqubit{~~~\Ucg~~~}{dd}\w & |\lambda\>\w \\
|q\> & \gspace{~~~\Ucg~~~}\w & |\lambda'\>\w\\
|i\> & \gspace{~~~\Ucg~~~}\w & |q\>\w\\
}

\caption{Schematic of the Clebsch-Gordan transform.  Equivalently, we
  could replace either the $\lambda$ output or the $\lambda'$ output
  with $j$.}\label{fig:cg}
\end{centering}
\end{figure}
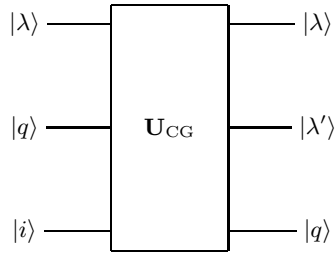

\subsection{Constructing the Schur Transform from Clebsch-Gordan
Transforms}\label{sec:schurcg}

We now describe how to construct the Schur transform out of a
series of Clebsch-Gordan transforms.  Suppose we start with an input
vector $\ket{i_1,\ldots,i_n} \in (\bbC^d)^{\ot n}$, corresponding to
the $\cU_d$-representation $(\cQ_{(1)}^d)^{\ot n}$.  According to
Schur duality (\eq{schur-decomp}), to perform the Schur transform it
suffices to
decompose $(\cQ_{(1)}^d)^{\ot n}$ into $\cU_d$-irreps.  This is
because Schur duality means that the multiplicity space of
$\cQ_\lambda^d$ must be isomorphic to $\cP_\lambda$.  In other words,
if we show that
\be (\cQ_{(1)}^d)^{\ot n} \stackrel{\cU_d}{\cong}
\bigoplus_{\lambda\in\bbZ_{++}^d}
\cQ_\lambda^d \ot \cP_\lambda',
\label{eq:pseudo-schur-decomp}\ee
then we must have $\cP_\lambda' \stackrel{\cS_n}{\cong} \cP_\lambda$
when $\lambda\in\cI_{d,n}$ and $\cP_\lambda' = \{0\}$ otherwise.

\begin{figure}[ht]
\begin{centering}
\includegraphics{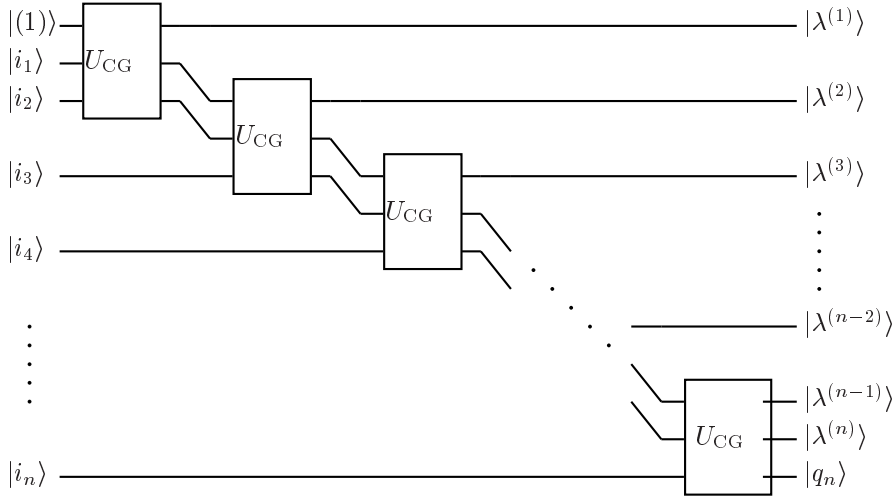}
\caption{Cascading Clebsch-Gordan transforms to produce the Schur transform.
Not shown are any ancilla inputs to the Clebsch-Gordan transforms.  The
structure of inputs and outputs of the Clebsch-Gordan transforms are the same
as in \fig{cg}.} \label{fig:cascade} 
\end{centering}
\end{figure}

To perform the $\cU_d$-irrep decomposition of
\eq{pseudo-schur-decomp}, we simply combine each of
$\ket{i_1},\ldots,\ket{i_n}$
using the CG transform, one at a time.  We start by inputting
$\ket{\lambda^{(1)}}=\ket{(1)}$, $\ket{i_1}$ and $\ket{i_2}$ into $\Ucg$
which outputs $\ket{\lambda^{(1)}}$ and a superposition of different values of
$\ket{\lambda^{(2)}}$ and $\ket{q_2}$.  Here $\lambda^{(2)}$ can be either
$(2,0)$ or $(1,1)$ and $\ket{q_2}\in Q_{\lambda^{(2)}}^d$.  Continuing, we
apply $\Ucg$ to $\ket{\lambda^{(2)}}\ket{q_2}\ket{i_3}$, and output a
superposition of vectors of the form
$\ket{\lambda^{(2)}}\ket{\lambda^{(3)}}\ket{q_3}$, with
$\lambda^{(3)}\in\cI_{d,3}$ and $\ket{q_3}\in Q_{\lambda^{(3)}}^d$.
Each time we are combining an arbitrary irrep $\lambda^{(k)}$ and
an associated basis vector $\ket{q_k}\in Q_{\lambda^{(k)}}^d$, together
with a vector from the defining irrep $\ket{i_{k+1}}$.  This is repeated
for $k=1,\ldots,n-1$ and the resulting circuit is depicted in
\fig{cascade}.

Finally, we are left with a superposition of states of the form
$\ket{\lambda^{(1)},\ldots,\lambda^{(n)}}\ket{q_n}$, where $\ket{q_n}\in
Q_{\lambda^{(n)}}^d$, $\lambda^{(k)}\in\cI_{d,k}$ and each $\lambda^{(k)}$ is
obtained by adding a single box to $\lambda^{(k-1)}$; i.e. $\lambda^{(k)} =
\lambda^{(k-1)} + e_{j_k}$ for some $j_k\in [d]$.  If we define
$\lambda=\lambda^{(n)}$ and $\ket{q}=\ket{q_n}$, then we have the
decomposition of
\eq{pseudo-schur-decomp} with $\cP_\lambda'$ spanned by the vectors
$\ket{\lambda^{(1)},\ldots, \lambda^{(n-1)}}$ satisfying the
constraints described above.  But this is
precisely the Young-Yamanouchi basis $P_\lambda$ that we have defined
in \sect{gz}!  Since the first $k$ qudits transform under
$\cU_d$ according to $\cQ_{\lambda^{(k)}}^d$, Schur duality implies that
they also transform under $\cS_k$ according to $\cP_{\lambda^{(k)}}$.
Thus we set $\ket{p}=\ket{\lambda^{(1)},\ldots,\lambda^{(n-1)}}$ (optionally
compressing to $\lceil\log|P_\lambda|\rceil$ qubits using the
techniques described in the last section) and obtain the desired
$\ket{\lambda}\ket{q}\ket{p}$.  As a check on this result, note that
each $\lambda^{(k)}$ is invariant under $\bQ(\cU_d)$ since $U^{\ot n}$
acts on the first $k$ qubits simply as $U^{\ot k}$.

If we choose not to perform the
$\poly(n)$ steps to optimally compress $\ket{\lambda^{(1)},\ldots,\lambda^{(n-1)}}$,
we could instead have our circuit output the equivalent
$\ket{j_1,\ldots,j_{n-1}}$, which requires only $n\log d$ qubits and
asymptotically no extra running time.

We can now appreciate the similarity between the $\cU_d$ CG ``add a
box'' prescription and the $\cS_{n-1}\subset \cS_n$ branching rule of
``remove a box.''  Schur duality implies that the representations
$\cQ_{\lambda'}^d$ that are obtained by decomposing $\cQ_\lambda^d \ot
\cQ_{(1)}^d$ are the same as the $\cS_n$-irreps $\cP_{\lambda'}$ that
include $\cP_\lambda$ when restricted to $\cS_{n-1}$.

Define $T_{\text{CG}}(n,d,\eps)$ to be the time complexity (in terms
of number of gates) of performing a single $\cU_d$ CG transform to
accuracy $\eps$ on Young diagrams with $\leq n$ boxes.  Then the total
complexity for the Schur transform is $n\cdot
(T_{\text{CG}}(n,d,\eps/n)+O(1))$, possibly plus a $\poly(n)$ factor
for compressing the $\cP_\lambda$ register to
$\lceil\log\dim\cP_\lambda\rceil$ qubits (as is required for
applications such as data compression and entanglement concentration,
cf.~\sect{schur-qit-apps}).  In the next section we will show that
$T_{\text{CG}}(n,d,\eps)$ is $\poly(\log n,d,\log 1/\eps)$, but first
we give a step-by-step description of the algorithm for the Schur
transform.

\begin{tabbing}
~~~~ \= ~~~ \= ~~~ \= ~~~ \= ~~~\= \kill
\+{\bf Algorithm: Schur transform (plus optional compression)} \\
{\bf Inputs:} (1) Classical registers $d$ and $n$.
 (2) An $n$ qudit quantum register
$\ket{i_1,\ldots,i_n}$.\\ 
{\bf Outputs:} Quantum registers $\ket{\lambda}\ket{q}\ket{p}$, with
$\lambda\in\cI_{d,n}$, $q\in Q_\lambda^d$ and $p\in P_\lambda$.\\
\pushtabs {\bf Runtime:} \=
$n\cdot (T_{\text{CG}}(n,d,\eps/n)+O(1))$ to achieve
accuracy $\eps$.\\
\> (Optionally plus $\poly(n)$ to compress the $\cP_\lambda$ register
to $\lceil \log\dim\cP_\lambda\rceil$ qubits.)\\\poptabs
{\bf Procedure:}\\
{\bf 1.}\> Initialize $\ket{\lambda^{(1)}}:=\ket{(1)}$ and
$\ket{q_1}=\ket{i_1}$.\\
{\bf 2.}\> For $k=1,\ldots,n-1$:\\
{\bf 3.}\>\> Apply $\Ucg$ to
$\ket{\lambda^{(k)}}\ket{q_k}\ket{i_{k+1}}$ to obtain output
$\ket{j_k}\ket{\lambda^{(k+1)}}\ket{q_{k+1}}$, where
$\lambda^{(k+1)}=\lambda^{(k)} + e_{j_k}$.\\
{\bf 4.}\> Output  $\ket{\lambda}:=\ket{\lambda^{(n)}}$,
$\ket{q}:=\ket{q_n}$ and $\ket{p}:=\ket{j_1,\ldots,j_{n-1}}$.\\
{\bf 5.}\> (Optionally use \eq{Plambda-relabel} to reversibly map
$\ket{j_1,\ldots,j_{n-1}}$ to an integer $p\in [\dim\cP_\lambda]$.)
\end{tabbing}

This algorithm will be made efficient in the next
section, where we efficiently construct the CG transform for $\cU_d$,
proving that $T_{\text{CG}}(n,d,\eps) = \poly(\log n, d, \log
1/\eps)$.

\section{Efficient circuits for the Clebsch-Gordan transform}
\label{sec:cg-construct}
We now turn to the actual construction of the circuit for the Clebsch-Gordan
transform described in \sect{cg-def}.  To get a feel for the what
will be necessary, we start by giving a circuit for the CG transform
that is efficient when $d$ is constant; i.e. it has complexity
$n^{O(d^2)}$, which is $\poly(n)$ for any constant value of $d$.

First recall that $\dim\cQ_\lambda^d \leq (n+1)^{d^2}$.  Thus, controlled on
$\lambda$, we want to construct a unitary transform on a $D$-dimensional system
for $D=\max_{\lambda\in\cI_{d,n}}\dim\cQ_\lambda^d = \poly(n)$.  There are
classical algorithms\cite{Louck:70a} to compute matrix elements of $\Ucg$ to an
accuracy $\epsilon_1$ in time $\poly(D)\poly\log(1/\epsilon_1)$.  Once we have
calculated all the relevant matrix elements (of which there are only
polynomially many), we can (again in time $\poly(D)\poly\log(1/\epsilon)$)
decompose $\Ucg$ into $D^2\poly\log(D)$ elementary one and two-qubit
operations\cite{Shende:04a,Reck:94a,Barenco:95a,Nielsen:00a}.  These can in
turn be approximated to accuracy $\epsilon_2$ by 
products of unitary operators from a fixed finite set (such as Clifford
operators and a $\pi/8$ rotation) with a further overhead of
$\poly\log(1/\epsilon_2)$\cite{Dawson05,Kitaev:02a}.  We can either assume
the relevant classical computations (such as decomposing the $D\times D$ matrix
into elementary gates) are performed coherently on a quantum computer, or as
part of a polynomial-time classical Turing machine which outputs the quantum
circuit. In any case, the total complexity is $\poly(n,\log 1/\epsilon)$ if the
desired final accuracy is $\epsilon$ and $d$ is held constant.

The goal of this section is to reduce this running time to $\poly(n,
d, \log(1/\epsilon))$; in fact, we will achieve circuits of size
$\poly(d,\log n, \log (1/\epsilon))$.  To do so, we will reduce the
$\cU_d$ CG transform to two components; first, a $\cU_{d-1}$ CG
transform, and second, a $d\times d$ unitary matrix whose entries can
be computed classically in $\poly(d,\log
n,1/\epsilon)$ steps.  After computing all $d^2$ entries, the second
component can then be implemented with $\poly(d, \log 1/\epsilon)$
gates according to the above arguments.

This reduction from the $\cU_d$ CG transform to the $\cU_{d-1}$ CG
transform is a special case of the Wigner-Eckart Theorem, which
we review in \sect{wigner}.  Then, following
\cite{Biedenharn:68a,Louck:70a}, we
use the Wigner-Eckart Theorem to give an efficient recursive
construction for $\Ucg$ in \sect{cg-recurse}.  Putting everything
together, we obtain a quantum circuit for the Schur transform that is
accurate to within $\epsilon$ and runs in time $n\cdot\poly(\log n, d, \log
1/\epsilon)$, optionally plus an additional $\poly(n)$ time to
compress the $\ket{p}$ register.

\subsection{The Wigner-Eckart Theorem and Clebsch-Gordan transform}
\label{sec:wigner}

In this section, we introduce the concept of an irreducible tensor operator,
which we use to state and prove the Wigner-Eckart Theorem.  Here we will
find that the CG transform is a key part of the Wigner-Eckart Theorem,
while in the next section we will turn this around and use the
Wigner-Eckart Theorem to give a recursive decomposition of the CG
transform.

Suppose $(\br_1,V_1)$ and $(\br_2,V_2)$ are representations of
$\cU_d$.  Recall that $\Hom(V_1,V_2)$ is a representation of $\cU_d$
under the map $T \ra \br_2(U) T \br_1(U)^{-1}$ for
$T\in\Hom(V_1,V_2)$.  If $\bmath{T}=\{T_1,T_2,\ldots\} \subset
\Hom(V_1,V_2)$ is a basis for a $\cU_d$-invariant subspace of
$\Hom(V_1,V_2)$, then we call $\bmath{T}$ a {\em tensor operator}.
Note that a tensor operator $\bT$ is a collection of operators
$\{T_i\}$ indexed by $i$, just as a tensor (or vector) is a collection
of scalars labeled by some index.  For example, the Pauli matrices
$\{\sigma_x,\sigma_y,\sigma_z\}\subset \Hom(\bbC^2,\bbC^2)$ comprise a
tensor operator, since conjugation by $\cU_2$ preserves the subspace
that they span.

Since $\Hom(V_1,V_2)$ is a representation of $\cU_d$, it can be
decomposed into irreps.  If $\bmath{T}$ is a basis for one of these
irreps, then we call it an {\em irreducible tensor operator}.  For
example, the Pauli matrices mentioned above comprise an irreducible
tensor operator, corresponding to the three-dimensional irrep
$\cQ_{(2)}^2$.  Formally, we say that $\bmath{T}^\nu =
\{T^\nu_{q_\nu}\}_{q_\nu \in Q_\nu^d}\subset \Hom(V_1,V_2)$ is an
irreducible tensor operator 
(corresponding to the irrep $\cQ_\nu^d$) if for all $U\in \cU_d$ we
have
\be 
\br_2(U) T^\nu_{q_\nu} \br_1(U)^{-1}
= \sum_{q'_\nu\in Q_\nu^d} \bra{q'_\nu}\bq_\nu^d(U)\ket{q_\nu}
T^\nu_{q'_\nu}. \label{eq:irred-tensor-op}\ee

Now assume that $V_1$ and $V_2$ are irreducible (say $V_1=\cQ_\mu^d$
and $V_2=\cQ_\lambda^d$), since if they are not, we could always
decompose $\Hom(V_1,V_2)$ into a direct sum of homomorphisms from an
irrep in $V_1$ to an irrep in $V_2$.  We can decompose
$\Hom(\cQ_\mu^d, \cQ_\lambda^d)$ into irreps using \eq{hom-decomp} and
the identity $\Hom(A,B) \cong A^* \ot B$ as follows:
\be\begin{split}
\Hom(\cQ_\mu^d, \cQ_\lambda^d) &\stackrel{\cU_d}{\cong}
\bigoplus_{\nu \in \bbZ_{++}^d} \cQ_\nu^d \ot
\Hom(\cQ_\nu^d, \Hom(\cQ_\mu^d, \cQ_\lambda^d))^{\cU_d} \\
&\stackrel{\cU_d}{\cong}
\bigoplus_{\nu \in \bbZ_{++}^d} \cQ_\nu^d \ot
\Hom(\cQ_\nu^d, (\cQ_\mu^d)^* \ot \cQ_\lambda^d)^{\cU_d} \\
&\stackrel{\cU_d}{\cong}
\bigoplus_{\nu \in \bbZ_{++}^d} \cQ_\nu^d \ot
\l((\cQ_\mu^d)^* \ot (\cQ_\nu^d)^* \ot \cQ_\lambda^d\r)^{\cU_d}\\
&\stackrel{\cU_d}{\cong}
\bigoplus_{\nu \in \bbZ_{++}^d} \cQ_\nu^d \ot
\Hom(\cQ_\mu^d \ot \cQ_\nu^d, \cQ_\lambda^d)^{\cU_d}
\end{split}\label{eq:hom-cg-decomp}\ee

Now consider a particular irreducible tensor operator $\bT^\nu
\subset \Hom(\cQ_\mu^d, \cQ_\lambda^d)$ with components
$T^\nu_{q_\nu}$ where $q_\nu$ ranges over $Q_\nu^d$.  We can define a
linear operator $\hat{T}: \cQ_\mu^d \ot \cQ_\nu^d \ra
\cQ_\lambda^d$ by letting
\be \hat{T} \ket{q_\mu}\ket{q_\nu} := T^\nu_{q_\nu}
\ket{q_\mu} \label{eq:Tnu-def}\ee
for all $q_\mu \in Q_\mu^d, q_\nu \in Q_\nu^d$ and extending it to the
rest of $\cQ_\mu^d \ot \cQ_\nu^d$ by linearity.
By construction, $\hat{T} \in \Hom(\cQ_\mu^d \ot \cQ_\nu^d,
\cQ_\lambda^d)$, but we claim that in addition $\hat{T}$ is
invariant under the action of $\cU_d$; i.e. that it lies in
$\Hom(\cQ_\mu^d \ot \cQ_\nu^d, \cQ_\lambda^d)^{\cU_d}$.  To see this,
apply \eqs{irred-tensor-op}{Tnu-def} to show that for any
$U\in\cU_d$, $q_\mu\in Q_\mu^d$ and $q_\nu\in Q_\nu^d$, we have
\be\begin{split}
\bq_\lambda^d(U) \hat{T}
\bigl[\bq_\mu^d(U)^{-1} \ot \bq_\nu^d(U)^{-1}\bigr]
\ket{q_\mu}\ket{q_\nu}
&= \sum_{q_\nu'\in Q_\nu^d}\bra{q'_\nu} \bq_\nu^d(U)^{-1}\ket{q_\nu}
\bq_\lambda^d(U) {T}^\nu_{q'_\nu} \bq_\mu^d(U)^{-1}
\ket{q_\mu} \\
&= \sum_{q_\nu',q_\nu''\in Q_\nu^d}
\bra{q''_\nu} \bq_\nu^d(U)\ket{q'_\nu}
\bra{q'_\nu} \bq_\nu^d(U)^{-1}\ket{q_\nu}
{T}^\nu_{q''_\nu} \ket{q_\mu}
\\ &= T^\nu_{q_\nu} \ket{q_\mu}
= \hat{T} \ket{q_\mu}\ket{q_\nu}.
\end{split}\ee

Now, fix an orthonormal basis for $\Hom(\cQ_\mu^d \ot \cQ_\nu^d,
\cQ_\lambda^d)^{\cU_d}$ and call it $M_{\mu,\nu}^\lambda$.  Then we
can expand $\hat{T}$ in this basis as
\be \hat{T} = \sum_{\alpha\in M_{\mu,\nu}^\lambda}
\hat{T}_\alpha \cdot \alpha,\ee
where the $\hat{T}_\alpha$ are scalars.  Thus
\be \bra{q_\lambda}T^\nu_{q_\nu}\ket{q_\mu}
= \sum_{\alpha\in M_{\mu,\nu}^\lambda}
\hat{T}_{\alpha}
\bra{q_\lambda}\alpha\ket{q_\mu,q_\nu}.
\label{eq:Talpha-def}\ee
This last expression $\bra{q_\lambda}\alpha\ket{q_\mu,q_\nu}$
bears a striking resemblance to the CG transform.  Indeed, note that
the multiplicity space $\Hom(\cQ_\lambda^d, \cQ_\mu^d \ot
\cQ_\nu^d)^{\cU_d}$ from \eq{cg-decomp} is the dual of $\Hom(\cQ_\mu^d
\ot \cQ_\nu^d, \cQ_\lambda^d)^{\cU_d}$ (which contains $\alpha$),
meaning that we can map between the two by taking the transpose.
In fact, taking the conjugate transpose of \eq{Ucg-dag} gives
$\bra{q_\lambda}\alpha = \bra{q_\lambda,\alpha^\dag}\Ucg^{\mu,\nu}$.  Thus
\be \bra{q_\lambda}\alpha\ket{q_\mu,q_\nu}
= \bra{q_\lambda, \alpha^\dag}\Ucg^{\mu,\nu}
\ket{q_\mu,q_\nu}.\ee

The arguments in the last few paragraphs constitute a proof of the
Wigner-Eckart theorem\cite{Messiah62}, which is
stated as follows:\\
\begin{theorem}[Wigner-Eckart]
\sloppypar{For any irreducible tensor operator
 $\bT^\nu = \{T^\nu_{q_\nu}\}_{q_\nu\in Q_\nu^d} \subset
 \Hom(\cQ_\mu^d,\cQ_\lambda^d)$,  there
exist $\hat{T}_\alpha\in\bbC$ for each $\alpha\in
M_{\mu,\nu}^\lambda$ such that for all $\ket{q_\mu}\in \cQ_\mu^d$,
$\ket{q_\nu}\in \cQ_\nu^d$ and $\ket{q_\lambda}\in \cQ_\lambda^d$:}
\be \bra{q_\lambda}T^\nu_{q_\nu} \ket{q_\mu}
= \sum_{\alpha\in M_{\mu,\nu}^\lambda}
\hat{T}_\alpha
\bra{q_\lambda,\alpha^\dag}\Ucg^{\mu,\nu}\ket{q_\mu,q_\nu}.\ee
\end{theorem}

Thus, the action of tensor operators can be related to a component
$\hat{T}_\alpha$ that is invariant under $\cU_d$ and a component
that is equivalent to the CG transform.  We will use this in the next
section to derive an efficient quantum circuit for the CG transform.

\subsection{A recursive construction of the Clebsch-Gordan transform}
\label{sec:cg-recurse}
In this section we show how the $\cU_d$ CG transform (which here we call
$\Ucg^{[d]}$) can be efficiently reduced to the $\cU_{d-1}$ CG
transform (which we call $\Ucg^{[d-1]}$).  Our strategy, following
\cite{Biedenharn:68a}, will be to express $\Ucg^{[d]}$ in terms of
$\cU_{d-1}$ tensor operators and then use the Wigner-Eckart Theorem to
express it in terms of $\Ucg^{[d-1]}$.  After we have explained this as a
relation among operators, we describe a quantum circuit for
$\Ucg^{[d]}$ that uses $\Ucg^{[d-1]}$ as a subroutine.

First, we express $\Ucg^{[d]}$ as a $\cU_d$ tensor operator.  For
$\mu\in\bbZ_{++}^d$, $\ket{q}\in Q_\mu^d$ and $i\in [d]$, we can expand
$\Ucg^{[d]} \ket{\mu}\ket{q}\ket{i}$ as
\be \Ucg^{[d]} \ket{\mu}\ket{q}\ket{i}
= \ket{\mu}
\sum_{\substack{j\in[d]\st\\ \mu + e_j\in\bbZ_{++}^d}}
\sum_{q'\in Q_{\mu+e_j}^d}
C^{\mu,j}_{q,i,q'}\ket{\mu + e_j}\ket{q'}.\ee
for some coefficients $C^{\mu,j}_{q,i,q'}\in \bbC$.
Now define operators $T^{\mu,j}_i: \cQ_\mu^d \ra \cQ_{\mu+e_j}^d$ by
\be T^{\mu,j}_i = \sum_{q\in Q_\mu^d}\;
\sum_{q'\in Q_{\mu+e_j}^d}
C^{\mu,j}_{q,i,q'} \ket{q'}\bra{q},\ee
so that $\Ucg^{[d]}$ decomposes as
\be \Ucg^{[d]} \ket{\mu}\ket{q}\ket{i}
= \ket{\mu}
\sum_{\substack{j\in[d]\st\\ \mu + e_j\in\bbZ_{++}^d}}
\ket{\mu + e_j}T^{\mu,j}_i \ket{q}.
\label{eq:CG-to-tensor}\ee
Thus $\Ucg^{[d]}$ can be understood in terms of the maps
$T_i^{\mu,j}$, which are irreducible tensor operators in
$\Hom(\cQ_\mu^d, \cQ_{\mu+e_j}^d)$  corresponding to the irrep
$\cQ_{(1)}^d$.  (This is unlike the notation of the last section in
which the superscript denoted the irrep corresponding to the tensor
operator.)

The plan for the rest of the section is to decompose the $T_i^{\mu,j}$
operators under the action of $\cU_{d-1}$, so that we can apply the
Wigner-Eckart theorem.  This involves decomposing three different
$\cU_d$ irreps into $\cU_{d-1}$ irreps: the input space $\cQ_\mu^d$,
the output space $\cQ_{\mu+e_j}^d$ and the space $\cQ_{(1)}^d$
corresponding to the subscript $i$.  Once we have done so,
the Wigner-Eckart Theorem gives an expression for $T_i^{\mu,j}$ (and
hence for $\Ucg^{[d]}$) in terms of $\Ucg^{[d-1]}$ and a small number
of coefficients, known as {\em reduced Wigner coefficients}.  These
coefficients can be readily calculated, and in the next section we
cite a formula from \cite{Biedenharn:68a} for doing so.

First, we examine the decomposition of $\cQ_{(1)}^d$, the $\cU_d$-irrep
according to which the $T_i^{\mu,j}$ transform.  Recall that
$\cQ_{(1)}^d \stackrel{\cU_{d-1}}{\cong} 
\cQ_{(0)}^{d-1} \oplus \cQ_{(1)}^{d-1}$.  In terms of the tensor
operator we have defined, this means that $T^{\mu,j}_d$ is an
irreducible $\cU_{d-1}$ tensor operator corresponding to the trivial
irrep $\cQ_{(0)}^{d-1}$ and $\{T^{\mu,j}_1, \ldots, T^{\mu,j}_{d-1}\}$
comprise an irreducible $\cU_{d-1}$ tensor operator corresponding to
the defining irrep $\cQ_{(1)}^{d-1}$.

Next, we would like to decompose $\Hom(\cQ_\mu^d, \cQ_{\mu+e_j}^d)$
into maps between irreps of $\cU_{d-1}$.  This is 
 slightly
more complicated, but can be derived from the $\cU_{d-1}\subset
\cU_d$ branching rule introduced in \sect{gz-yy}.  Recall that
$\cQ_\mu^d \stackrel{\cU_{d-1}}{\cong} \bigoplus_{\mu'\interlaces \mu}
\cQ_{\mu'}^{d-1}$, and similarly $\cQ_{\mu+e_j}^d
\stackrel{\cU_{d-1}}{\cong} \bigoplus_{\mu''\interlaces \mu+e_j}
\cQ_{\mu''}^{d-1}$.  This is the moment that we anticipated in
\sect{gz-yy} when we chose our set of basis vectors
$Q_\mu^d$ to respect these decompositions.   As a result,  a vector
$\ket{q}\in Q_\mu^d$ can be expanded as
$q = (q_{d-1},q_{d-2},\ldots,q_1) = (\mu',q_{(d-2)})$ with
$q_{d-1}=\mu'\in\bbZ_{++}^{d-1}$, $\mu'\interlaces \mu$ and $\ket{q_{(d-2)}}
= \ket{q_{d-2},\ldots,q_1}\in Q_{\mu'}^{d-1}$.  In other words, we
will separate vectors in $Q_\mu^d$ into a $\cU_{d-1}$ irrep label
$\mu'\in\bbZ_{++}^{d-1}$ and a basis vector from $\cQ_{\mu'}^{d-1}$.

This describes how to decompose the spaces $\cQ_\mu^d$ and
$\cQ_{\mu+e_j}^d$.  To extend this to decomposition of
$\Hom(\cQ_\mu^d, \cQ_{\mu+e_j}^d)$, we use the 
canonical isomorphism $\Hom(\bigoplus_x A_x, \bigoplus_y B_y) \cong
\bigoplus_{x,y} \Hom(A_x,B_y)$, which holds for any sets of vector
spaces $\{A_x\}$ and $\{B_y\}$.  Thus
\begin{subequations}\label{eq:wigner-op-decomp}
\be \Hom(\cQ_\mu^d, \cQ_{\mu+e_j}^d)  \stackrel{\cU_{d-1}}{\cong}
\bigoplus_{\mu'\interlaces\mu} \;\; \bigoplus_{\mu''\interlaces\mu+e_j}
\Hom(\cQ_{\mu'}^{d-1}, \cQ_{\mu''}^{d-1}).
\label{eq:wigner-op-decomp1}\ee
Sometimes we will find it convenient to denote the
$\cQ_{\mu'}^{d-1}$ subspace of $\cQ_\mu^d$ by
$\cQ_{\mu'}^{d-1} \subset \cQ_\mu^d$, so that \eq{wigner-op-decomp1}
becomes
\be \Hom(\cQ_\mu^d, \cQ_{\mu+e_j}^d)  \stackrel{\cU_{d-1}}{\cong}
\bigoplus_{\mu'\interlaces\mu}\;\; \bigoplus_{\mu''\interlaces\mu+e_j}
\Hom(\cQ_{\mu'}^{d-1}\subset\cQ_\mu^d,
\cQ_{\mu''}^{d-1}\subset\cQ_{\mu+e_j}^d).
\label{eq:wigner-op-decomp2}\ee
\end{subequations}

According to \eq{wigner-op-decomp} (either version), we can decompose
$T^{\mu,j}_i$ as 
\be T^{\mu,j}_i =
\sum_{\mu'\interlaces \mu}\;\;\sum_{\mu''\interlaces \mu+e_j}
\ket{\mu''}\bra{\mu'} \ot T^{\mu,j,\mu',\mu''}_i.
\label{eq:tensor-decomp}\ee
Here $T^{\mu,j,\mu',\mu''}_i \in
\Hom(\cQ_{\mu'}^{d-1}\subset\cQ_\mu^d,
\cQ_{\mu''}^{d-1}\subset\cQ_{\mu+e_j}^d)$ and we have implicitly
decomposed $\ket{q}\in Q_\mu^d$ into
$\ket{\mu'}\ket{q_{(d-2)}}$.

The next step is to decompose the representions in
\eq{wigner-op-decomp} into irreducible 
components.  In fact, we are not
interested in the entire space $\Hom(\cQ_{\mu'}^{d-1},
\cQ_{\mu''}^{d-1})$, but only the part that is equivalent to
$\cQ_{(1)}^{d-1}$ or $\cQ_{(0)}^{d-1}$, depending on whether
$i\in[d-1]$ or $i=d$ (since $T^{\mu,j,\mu',\mu''}_i$ transforms
according to $\cQ_{(1)}^{d-1}$ if $i\in\{1,\ldots,d-1\}$ and
according to $\cQ_{(0)}^{d-1}$ if $i=d$).  This knowledge of how
$T^{\mu,j,\mu',\mu''}_i$ transforms under $\cU_{d-1}$ will give us two
crucial simplifications: first, we can greatly reduce the range of
$\mu''$ for which $T^{\mu,j,\mu',\mu''}_i$ is nonzero, and second, we
can apply the Wigner-Eckart theorem to describe
$T^{\mu,j,\mu',\mu''}_i$ in terms of $\Ucg^{[d-1]}$.

The simplest case is $\cQ_{(0)}^{d-1}$, when $i=d$: according to
Schur's Lemma 
the invariant component of $\Hom(\cQ_{\mu'}^{d-1},
\cQ_{\mu''}^{d-1})$ is zero if $\mu'\neq\mu''$ and consists of the
matrices proportional to $I_{\cQ_{\mu'}^{d-1}}$ if $\mu'=\mu''$.  In
other words $T_d^{\mu,j,\mu',\mu''}=0$ unless $\mu'=\mu''$, in which
case $T_d^{\mu,j,\mu',\mu'} := \hat{T}^{\mu,j,\mu',0}
I_{\cQ_{\mu'}^{d-1}}$ for some scalar $\hat{T}^{\mu,j,\mu',0}$.  (The
final superscript 0 will later be convenient when we want a single
notation to encompass both the $i=d$ and the $i\in\{1,\ldots,d-1\}$
cases.)

The $\cQ_{(1)}^{d-1}$ case, which occurs when $i\in\{1,\ldots,d-1\}$, is more
interesting.  We will simplify the $T^{\mu,j,\mu',\mu''}_i$ operators
(for $i=1,\ldots,d-1$) in two stages: first using the branching rules
from \sect{gz-yy} to reduce the number of nonzero terms and then by
applying the Wigner-Eckart theorem to find an exact expression for
them. Begin by recalling from \eq{hom-cg-decomp} that the multiplicity of
$\cQ_{(1)}^{d-1}$ in the isotypic decomposition of
$\Hom(\cQ_{\mu'}^{d-1}, \cQ_{\mu''}^{d-1})$ is 
given by $\dim\Hom(\cQ_{\mu'}^{d-1} \ot \cQ_{(1)}^{d-1},
\cQ_{\mu''}^{d-1})^{\cU_{d-1}}$.  According to the $\cU_{d-1}$ CG
``add a box''  prescription (\eq{cg-add-one}), this is one if
 $\mu'\in\mu''-\Box$ 
 and zero otherwise.  Thus if $i\in [d-1]$, then
$T^{\mu,j,\mu',\mu''}_i$ is zero unless $\mu'' = \mu' + e_{j'}$
for some $j'\in [d-1]$.  Since we need not consider all possible
$\mu''$, we can define $T^{\mu,j,\mu',j'}_i := T^{\mu,j,\mu',\mu' +
e_{j'}}_i$.
This notation can be readily extended to cover the case when
$i=d$; define $e_0=0$, so that the only nonzero operators for
$i=d$ are of the form $T^{\mu,j,\mu',0}_d := T^{\mu,j,\mu',\mu'}_d =
\hat{T}^{\mu,j,\mu',0} I_{\cQ_{\mu'}^{d-1}}$.  Thus, we can 
replace \eq{tensor-decomp} with
\be T^{\mu,j}_i =
\sum_{\mu'\interlaces \mu}\;\sum_{j'=0}^{d-1}
\ket{\mu'+e_{j'}}\bra{\mu'} \ot T^{\mu,j,\mu',\mu'+e_{j'}}_i.
\label{eq:tensor-decomp2}\ee

Now we show how to apply the Wigner-Eckart theorem to the $i\in[d-1]$
case.  The operators $T^{\mu,j,\mu',j'}_i$
map $\cQ_{\mu'}^{d-1}$ to $\cQ_{\mu'+e_{j'}}^{d-1}$ and comprise an
irreducible $\cU_{d-1}$ tensor operator corresponding to the irrep
$\cQ_{(1)}^{d-1}$. This means
we can apply the Wigner-Eckart theorem and since the multiplicity of
$\cQ_{\mu'+e_{j'}}^{d-1}$ in $\cQ_{\mu'}^{d-1}\ot \cQ_{(1)}^{d-1}$ is
one, the sum over the multiplicity label $\alpha$ has only a single
term.  The theorem 
implies the existence of a set of scalars
$\hat{T}^{\mu,j,\mu',j'}$ such that for any $\ket{q}\in
Q_{\mu'}^{d-1}$ and $\ket{q'} \in Q_{\mu'+e_{j'}}^{d-1}$,
\be \bra{q'} T_i^{\mu,j,\mu',j'} \ket{q} =
\hat{T}^{\mu,j,\mu',j'}
\bra{\mu',\mu'+e_{j'},q'} \Ucg^{[d-1]} \ket{\mu',q,i}.
\label{eq:wigner-tensor-reduction}\ee
Sometimes the matrix elements of $\Ucg$ or
$T^{\mu,j,\mu',j'}_i$ are called {\em Wigner coefficients} and
the $\hat{T}^{\mu,j,\mu',j'}$ are known as {\em reduced Wigner
coefficients}.  

Let us now try to interpret these equations operationally.  
\eq{CG-to-tensor} reduces the $\cU_d$ CG transform to a $\cU_d$ tensor
operator, \eq{tensor-decomp2} decomposes this tensor operator into
$d^2$ different $\cU_{d-1}$ tensor operators (weighted by the
$\hat{T}^{\mu,j,\mu',j'}$ coefficients) and
\eq{wigner-tensor-reduction} turns this into a $\cU_{d-1}$ CG
transform followed by a $d\times d$ unitary matrix.  The coefficients
for this matrix are the $\hat{T}^{\mu,j,\mu',j'}$, which we will see
in the next section can be efficiently computed by conditioning on
$\mu$ and $\mu'$.

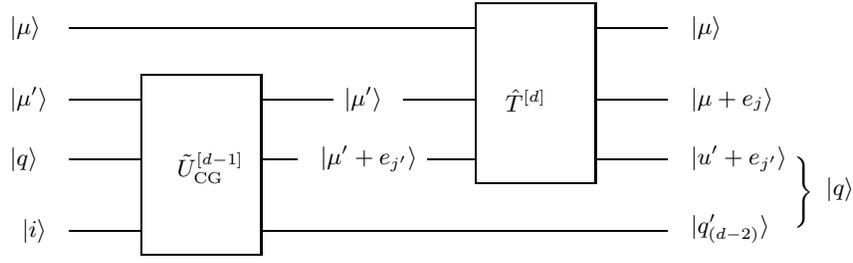
\begin{figure}[ht]
\begin{centering}
\setlength{\unitlength}{3947sp}%
\begin{picture}(5250,1599)(376,-1198)
\thinlines
{\put(1201,-1186){\framebox(750,1125){}}}
{\put(1201,-211){\line(-1, 0){450}}}
{\put(1201,-586){\line(-1, 0){450}}}
{\put(1201,-1036){\line(-1, 0){450}}}
{\put(2401,-211){\line(-1, 0){450}}}
{\put(2176,-586){\line(-1, 0){225}}}
{\put(4501,-1036){\line(-1, 0){2550}}}
{\put(3301,-586){\line(-1, 0){300}}}
{\put(3301,-211){\line(-1, 0){450}}}
{\put(3301,-736){\framebox(750,1125){}}}
{\put(4501,-211){\line(-1, 0){450}}}
{\put(4501,239){\line(-1, 0){450}}}
{\put(4501,-586){\line(-1, 0){450}}}
{\put(3301,239){\line(-1, 0){2550}}}
\put(1426,-736){\makebox(0,0)[lb]{$\tilde{U}_{\text{CG}}^{[d-1]}$}}
\put(3486,-286){\makebox(0,0)[lb]{$\hat{T}^{[d]}$}}
\put(376,-286){\makebox(0,0)[lb]{$|\mu'\>$}}
\put(4651,-661){\makebox(0,0)[lb]{$|u'+e_{j'}\>$}}
\put(4651,-1111){\makebox(0,0)[lb]{$|q'_{(d-2)}\>$}}
\put(4651,164){\makebox(0,0)[lb]{$|\mu\>$}}
\put(451,-1111){\makebox(0,0)[lb]{$|i\>$}}
\put(376,-661){\makebox(0,0)[lb]{$|q\>$}}
\put(2326,-661){\makebox(0,0)[lb]{$|\mu'+e_{j'}\>$}}
\put(2476,-286){\makebox(0,0)[lb]{$|\mu'\>$}}
\put(4651,-286){\makebox(0,0)[lb]{$|\mu+e_j\>$}}
\put(5300,-1020){\makebox(0,0)[lb]{$$\Bigg\}$$}}
\put(5500,-850){\makebox(0,0)[lb]{$|q\>$}}
\put(376,164){\makebox(0,0)[lb]{$|\mu\>$}}
\end{picture}
\caption{The $\cU_d$ CG transform, $\Ucg^{[d]}$, is decomposed into a
$\cU_{d-1}$ CG transform $\tilde{U}_{\text{CG}}^{[d-1]}$ (see 
\eq{tildeUcg-def}) and a reduced Wigner operator
$\hat{T}^{[d]}$.  In \fig{reduced-wigner} we show how to reduce
the reduced Wigner operator to a $d\times d$ matrix
conditioned on $\mu$ and $\mu'+e_{j'}$.
}
\label{fig:CG-decomp}
\end{centering}
\end{figure}

\begin{figure}[ht]
\begin{center}\leavevmode
\xymatrix@R=4pt@C=4pt{
{|\mu\>} & ~~~ & \gnqubit{~\hat{T}^{[d]}~}{dd}\ar@{-}[ll] &
*{~~~} & |\mu\>\ar@{-}[ll] \\
{|\mu'\>} & \nw & \gspace{~\hat{T}^{[d]}~}\w &\nw& |\mu+e_j\>\w\\
{|\mu'+e_{j'}\>} & \nw & \gspace{~\hat{T}^{[d]}~}\w
&\nw& |\mu'+e_{j'}\>\w
}
\xymatrix@R=4pt@C=4pt{
& |\mu\> & \nw & \nw & \b\w\ar@{-}[d] & \nw & \b\w\ar@{-}[d] & |\mu\>\w \\
*{~\Huge\mbox{$\cong$}~} & |\mu'\> & *+={\oplus}\w & *+{|e_{j'}\>}\w & 
\op{\hat{T}^{[d]}_{\mu,\mu'}}\w & *+{|e_j\>}\w & *+={\oplus}\w
& |\mu+e_j\>\w\\
& |\mu'+e_{j'}\> & \b\w\ar@{-}[u] & \nw & \b\w\ar@{-}[u] & \nw & \nw &
|\mu'+e_{j'}\>\w
}
\caption{The reduced Wigner transform $\hat{T}^{[d]}$ can be
expressed as a $d\times d$ rotation whose coefficients are controlled
by $\mu$ and $\mu'+e_{j'}$.}
\label{fig:reduced-wigner}
\end{center}
\end{figure}
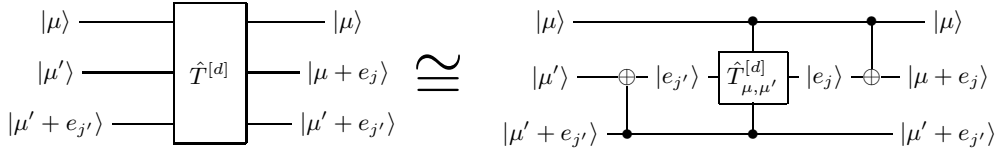

Now we spell this recursion out in more detail.  Suppose we wish to
apply $\Ucg^{[d]}$ to
$\ket{\mu}\ket{q}\ket{i}=\ket{\mu}\ket{\mu'}\ket{q_{(d-2)}}\ket{i}$,
for some $i\in\{1,\ldots,d-1\}$.  Then 
\eq{wigner-tensor-reduction} indicates that we should first apply
$\Ucg^{[d-1]}$ to $\ket{\mu'}\ket{q_{(d-2)}}\ket{i}$ to obtain output
that is a superposition over states
$\ket{\mu'+e_{j'}}\ket{j'}\ket{q'_{(d-2)}}$ for $j'\in\{1,\ldots,d-1\}$ and
$\ket{q'_{(d-2)}}\in Q_{\mu'+e_{j'}}^{d-1}$.  Then, controlled by
$\mu$ and $\mu'$, we want to map the $(d-1)$-dimensional $\ket{j'}$
register into the $d$-dimensional $\ket{j}$ register, which will then
tell us the output irrep $\cQ_{\mu+e_j}^d$.
According to \eq{wigner-tensor-reduction}, the coefficients of this
$d\times(d-1)$ matrix are given by the reduced Wigner coefficients
$\hat{T}^{\mu,j,\mu',j'}$,
 so we will denote the overall matrix
$\hat{T}^{[d]}_{\mu,\mu'} := \sum_{j,j'}
\hat{T}^{\mu,j,\mu'+e_{j'},j'} \ket{j}\!\bra{j'}$.\footnote{The reason
 why $\mu'+e_{j'}$ appears in the superscript rather than $\mu'$ is
 that after applying $\hat{T}^{[d]}_{\mu,\mu'}$ we want to keep a
 record of $\mu'+e_{j'}$ rather than of $\mu'$.  This is further
 illustrated in \fig{reduced-wigner}.}
  The resulting circuit is depicted in \fig{CG-decomp}: a $\cU_{d-1}$
CG transform is followed by the $\hat{T}^{[d]}$ operator, which is
defined to be
\be\hat{T}^{[d]} = \sum_{\mu'\interlaces\mu}\sum_{j,j'}
\hat{T}^{\mu,j,\mu',j'} \oprod{\mu} \ot \ket{\mu+e_j}\bra{\mu'}
\ot \oprod{\mu' + e_{j'}}.
\label{eq:red-wig-def}\ee
Then \fig{reduced-wigner} shows how $\hat{T}^{[d]}$ can be expressed
as a $d\times (d-1)$ matrix $\hat{T}^{[d]}_{\mu,\mu'}$ that is
controlled by $\mu$ and $\mu'$.  In fact, once we consider the $i=d$
case in the next paragraph, we will find that
$\hat{T}^{[d]}_{\mu,\mu'}$ is actually a $d\times d$ unitary matrix.
In the next section, we will then show how the individual reduced
Wigner coefficients $\hat{T}^{\mu,j,\mu',j'}$ can be efficiently
computed, so that ultimately $\hat{T}^{[d]}_{\mu,\mu'}$ can be
implemented in time $\poly(d,\log1/\eps)$.

Now we turn to the case of $i=d$.   The circuit is much simpler, but
we also need to explain how it 
works in coherent superposition with the $i\in[d-1]$ case.  Since
$i=d$ corresponds to the trivial representation of $\cU_{d-1}$, the
$\Ucg^{[d-1]}$ operation is not performed.  Instead, $\ket{\mu'}$ and
$\ket{q_{(d-2)}}$ are left untouched and the $\ket{i}=\ket{d}$
register is relabeled as a $\ket{j'}=\ket{0}$ register.  We can
combine this relabeling operation with $\Ucg^{[d-1]}$ in the
$i\in[d-1]$ case by defining
\be \tilde{U}_{\text{CG}}^{[d-1]} :=
\l(\ket{0}\bra{d} \ot \sum_{\mu'\in\bbZ_{++}^{d-1}} \oprod{\mu'}\r)
\ot I_{\cQ_{\mu'}^{d-1}} + \Ucg^{[d-1]}.
\label{eq:tildeUcg-def}\ee
This ends up mapping $i\in\{1,\ldots,d\}$ to $j'\in\{0,\ldots,d-1\}$
while mapping $\cQ_{\mu'}^{d-1}$ to $\cQ_{\mu'+e_{j'}}^{d-1}$.
Now we can interpret the sum on $j'$ in the above definitions of
$\hat{T}^{[d]}$ and $\hat{T}^{[d]}_{\mu,\mu'}$ as ranging over
$\{0,\ldots,d-1\}$, so that $\hat{T}^{[d]}_{\mu,\mu'}$ is a $d\times
d$ unitary matrix.  We thus obtain the circuit in \fig{CG-decomp} with
the implementation of $\hat{T}^{[d]}$ depicted in \fig{reduced-wigner}.

We have now reduced the problem of performing the CG transform
$\Ucg^{[d]}$ to the problem of computing reduced Wigner coefficients
$\hat{T}^{\mu,j,\mu',j'}$.

\subsection{Efficient Circuit for the Reduced Wigner Operator}

The method of Biedenharn and Louck\cite{Biedenharn:68a} allows us to
compute reduced Wigner coefficients for the cases we are interested
in.  This will allow us to construct an efficient circuit to implement the
controlled-$\hat{T}$ operator to accuracy $\epsilon$ using an overhead which
scales like $\poly(\log n, d, \log(\epsilon^{-1}))$.

To compute $\hat{T}^{\mu,j,\mu',j'}$, we first introduce the vectors
$\tilde{\mu} := \mu + \sum_{j=1}^d (d-j)e_j$ and
$\tilde{\mu}' := \mu' + \sum_{j=1}^{d-1} (d-1-j)e_j$.  
Also
define $S(j-j')$ to be 1 if $j\geq j'$ and $-1$ if $j<j'$.  Then
according to Eq.~(38) in
Ref~\cite{Biedenharn:68a},
\begin{equation}
\hat{T}^{\mu,j,\mu',j'} = 
\l\{\begin{array}{ll}
S(j-j')
\l[\frac{\prod_{s\in [d-1]\backslash j} 
(\tilde{\mu}_j - \tilde{\mu}'_s)
\prod_{t\in [d]\backslash j'} 
(\tilde{\mu}'_{j'} - \tilde{\mu}_t + 1)
}{
\prod_{s\in [d]\backslash j}
(\tilde{\mu}'_j - \tilde{\mu}'_s)
\prod_{t\in [d-1]\backslash j'} 
(\tilde{\mu}'_{j'} - \tilde{\mu}'_t + 1)
}\r]^{\frac{1}{2}}
& \mbox{ if $j'\in\{1,\ldots,d-1\}$.}\\
S(j-d)
\l[\frac{\prod_{s\in [d-1]\backslash j} 
(\tilde{\mu}_j - \tilde{\mu}'_s)
}{
\prod_{s\in [d]\backslash j}
(\tilde{\mu}'_j - \tilde{\mu}'_s)
}\r]^{\frac{1}{2}}
& \mbox{ if $j'=0$.}
\end{array}\r. 
\end{equation}
The elements of the partitions here are of size $O(n)$, so
the total computation necessary is $\poly(d, \log n)$.  Now how do we
implement the $\hat{T}^{[d]}$ transform given this
expression?

As in the introduction to this section, note that any unitary gate of
dimension $d$ can be 
implemented using a number of two qubit gates polynomial in
$d$\cite{Reck:94a,Barenco:95a,Nielsen:00a}.  The method of this
construction is to take a unitary gate of dimension $d$ with {\em
known} matrix elements and then convert this into a series of unitary
gates which act non-trivially only on two states.  These two state
gates can then be constructed using the methods described in
\cite{Barenco:95a}.  In order to modify this for our work, we
calculate, to the specified accuracy $\epsilon$, the elements of
the $\hat{T}^{[d]}$ operator, conditional on the $\mu$ and
$\mu'+e_{j'}$ inputs, perform the decomposition into two qubit gates
as described in \cite{Reck:94a,Barenco:95a} {\em online}, and then,
conditional on this calculation perform the appropriate controlled
two-qubit gates onto the space where $\hat{T}^{[d]}$ will act.
Finally this classical computation must be undone to reset any garbage
bits created during the classical computation.   To produce
an accuracy $\epsilon$ we need a classical computation of size ${\rm
poly}(\log(1/\epsilon))$ since we can perform the appropriate
controlled rotations with bitwise accuracy.

Putting everything together as depicted in figures \ref{fig:CG-decomp}
and \ref{fig:reduced-wigner} gives a $\poly(d,\log n,\log 1/\epsilon)$
algorithm to reduce $\Ucg^{[d]}$ to $\Ucg^{[d-1]}$.  Naturally this
can be applied $d$ times to yield a $\poly(d,\log n,\log 1/\epsilon)$
algorithm for $\Ucg^{[d]}$.  (We can end the recursion either at
$d=2$, using the construction in \cite{Bacon:04a}, or at $d=1$, where
the CG transform simply consists of the map $\mu\ra \mu+1$ for
$\mu\in\bbZ$, or even at $d=0$, where the CG transform is completely
trivial.)  We summarize the CG algorithm as follows.

\begin{tabbing}
~~~~ \= ~~~~ \= ~~~ \= ~~~ \= ~~~\= \kill
\+{\bf Algorithm: Clebsch-Gordan transform} \\
\parbox[t]{5.6in}{{\bf Inputs:} (1) Classical registers $d$ and
$n$. (2) Quantum 
registers $\ket{\lambda}$ (in any superposition over different
$\lambda\in\cI_{d,n}$), $\ket{q}\in \cQ_\lambda^d$ (expressed as a
superposition of GZ basis elements) and $\ket{i}\in\bbC^d$.}\\
\parbox[t]{5.6in}{{\bf Outputs:} (1) Quantum registers $\ket{\lambda}$
(equal to the 
input), $\ket{j}\in\bbC^d$ (satisfying $\lambda+e_j\in\cI_{d,n+1}$)
and $\ket{q'}\in\cQ_{\lambda+e_j}^d$.}\\
{\bf Runtime:} $d^3\poly(\log n,\log 1/\eps)$ to achieve accuracy
$\eps$.\\
{\bf Procedure:}\\
{\bf 1.} \> If $d=1$\\
{\bf 2.} \> Then output $\ket{j}:=\ket{i}=\ket{1}$ and
$\ket{q'}:=\ket{q}=\ket{1}$ (i.e. do nothing).\\
{\bf 3.} \> Else\\
{\bf 4.} \> \> Unpack $\ket{q}$ into
$\ket{\mu'}\ket{q_{(d-2)}}$, 
such that $\mu'\in\cI_{d,m}$, $m\leq n$, $\mu'\interlaces\mu$ and
$\ket{q_{(d-2)}}\in\cQ_{\mu'}^{d-1}$.\\ 
{\bf 5.} \> \> If $i<d$\\
{\bf 6.} \> \> \parbox[t]{5in}{Then perform the CG transform with inputs
$(d-1,m,\ket{\mu'},\ket{q_{(d-2)}},\ket{i})$ and outputs
$(\ket{\mu'},\ket{j'},\ket{q'_{(d-2)}})$.}\\
{\bf 7.} \> \> Else (if $i=d$)\\
{\bf 8.} \> \> \> \parbox[t]{5.5in}{Replace $\ket{i}=\ket{d}$ with
$\ket{j'}:=\ket{0}$ and set $\ket{q'_{(d-2)}}:=\ket{q'_{(d-2)}}$.}\\
{\bf 9.} \> \> End. (Now $i\in\{1,\ldots,d\}$ has been replaced by
$j\in\{0,\ldots, d-1\}$.)\\
{\bf 10.} \> \> Map $\ket{\mu'}\ket{j'}$ to
$\ket{\mu'+e_{j'}}\ket{j'}$.\\
{\bf 11.} \> \> \parbox{5.5in}{Conditioned on $\mu$ and $\mu'+e_j'$,
calculate the gate sequence necessary to implement $\hat{T}^{[d]}$,
which inputs $\ket{j'}$ and outputs $\ket{j}$.}\\
{\bf 12.} \> \> Execute this gate sequence, implementing
$\hat{T}^{[d]}$.\\
{\bf 13.} \> \> Undo the computation from {\bf 11}.\\
{\bf 14.} \> \> Combine $\ket{\mu'+e_{j'}}$ and $\ket{q'_{(d-2)}}$ to
form $\ket{q'}$.\\
{\bf 15.} \> End.
\end{tabbing}

Finally, in \sect{schur-circuit} we described
how $n$ CG transforms can be used to perform the Schur transform, so
that $\Usch$ can be implemented in time $n\cdot \poly(d,\log n,\log
1/\epsilon)$, optionally plus an additional $\poly(n)$ time to
compress the $\ket{p}$ register.

\section{Conclusion}

We have taken on the challenge of implementing a circuit which performs the
Schur transform.  This transform, used
ubiquitously\cite{Keyl:01a,Gill:02a,Vidal:99a,Hayashi:02a,Hayashi:02b,
Hayashi:02c,Hayashi:02d,Zanardi:97a,Knill:00a,Kempe:01a,Bacon:01a,Bartlett:03a}
in quantum information theory, represents an important new transformation for
quantum information science. The key ingredients in the construction of this
circuit were the relationship between Wigner operators and reduced Wigner
operators and an efficient classical algorithm for the calculation of the
matrix elements of the reduced Wigner operators.  This extends our construction
from \cite{Bacon:04a} where we constructed the Schur transform for $n$ qubits
($d=2$).  Our construction has a running time which is polynomial in dimension,
$d$, number of qudits, $n$, and accuracy, $\log(1/\epsilon$. We have thus made
practical the large set of quantum information protocols whose computational
efficiency has, prior to our work, been uncertain.

For some applications, it is not necessary to perform the full Schur transform,
but instead to only be able to perform a projective measurement onto the
different Schur subspaces. In part II, we consider a quantum circuit, based on
Kitaev's phase estimation algorithm\cite{Kitaev:95a}, for this task.  We
further generalize this algorithm to a circuit which is applicable to any
nonabelian finite group. Our algorithm is efficient if there exists an
efficient quantum circuit for the Fourier transform over this
group\cite{Beals:97a,Moore:03a} and represents an ideal way to efficiently deal
with situations where quantum states possess symmetries corresponding to some
finite group.  Further in part II we discuss relationships between the Schur
transform and the Fourier transform over the symmetric group.

Finally, we will conclude with some open problems suggested by our construction
of the Schur transform.  The first interesting question which arises from our
work is whether Clebsch-Gordan transforms for other groups can be efficiently
constructed.  We suspect that for many finite groups, even when dealing with
representations which are of dimension $d$, that their Clebsch-Gordan
transforms can be constructed using circuits of size polynomial in $\log(d)$.
Our intuition for this claim comes from the construction of quantum Fourier
transforms over finite groups\cite{Beals:97a,Moore:03a}.  A second question is
that while the Schur transform is used frequently in quantum information
theory, it has thus far not seen used in the field of quantum algorithms.
Kuperberg's\cite{Kuperberg:03a} subexponential algorithm for the dihedral
hidden subgroup problem makes use of the effect of the Clebsch-Gordan series
for the dihedral group. Does the Clebsch-Gordan series for ${\mathcal U}_d$
produce any similar speedup for the appropriately defined hidden subgroup
problem on ${\mathcal U}_d$?

\bigskip
{\em Acknowledgments:}
This work was partially funded by the NSF Institute for Quantum
Information under grant number EIA-0086048.  AWH acknowledges partial
support from the NSA and ARDA under ARO contract DAAD19-01-1-06.

\bibliographystyle{unsrt}
\bibliography{bigref}

\end{document}